\documentclass[acmsmall]{acmart}
\allowdisplaybreaks
\usepackage[ruled,vlined,linesnumbered]{algorithm2e}
\usepackage{graphicx,enumitem,booktabs,amsmath}
\usepackage{psfrag}
\usepackage{marginnote}
\usepackage{tikz}
\usepackage{listings}
\usepackage[tableposition=top]{caption}
\usepackage{subfigure}
\usepackage{color}
\usepackage{booktabs}
\usepackage{subfigure}
\usepackage{booktabs}
\usepackage[a-2b]{pdfx}
\usepackage{amsmath}

\usepackage{mathabx}

\usepackage[ruled, vlined, linesnumbered]{algorithm2e}
\usepackage{enumitem}
\usepackage[utf8]{inputenc}
\usepackage{latexsym}
\usepackage{xcolor,colortbl}
\usepackage{hhline}
\usepackage{cancel}
\usepackage{tikzscale}
\usepackage{soul} 
\usepackage{balance}

\usepackage{multirow}

\newcommand{\nnext}{\textup{\textsf{next}}}
\newcommand{\sskip}{\textup{\textsf{skip}}}
\newcommand{\remain}{\textup{\textsf{remain}}}
\newcommand{\random}{\textup{\textsf{rand}}}
\renewcommand{\paragraph}[1]{\smallskip \noindent{\bf {#1. }}}

\newcommand{\cnt}{\textup{\textsf{cnt}}}
\newcommand{\wcnt}{\Tilde{\textup{\textsf{cnt}}}} 
\newcommand{\nsum}{N}

\newcommand{\mul}{\textsf{feq}}
\newcommand{\wmul}{\Tilde{\textup{\textsf{feq}}}}

\newcommand{\nnull}{\textup{\textbf{null}}}

\newcommand{\OUT}{\textsf{OUT}}

\newcommand{\Q}{\mathcal{Q}}
\newcommand{\E}{\mathcal{E}}
\newcommand{\C}{\mathcal{C}}

\newcommand{\V}{\mathcal{V}}

\newcommand{\dom}{\mathrm{dom}}

\newcommand{\R}{\mathcal{R}}
\newcommand{\T}{\mathcal{T}}

\newcommand{\key}{\textsf{key}}
\newcommand{\pos}{\textsf{pos}}
\newcommand{\batchres}{\textsf{RSJoin}}
\newcommand{\batchresfk}{\textsf{RSJoin\textunderscore opt}}
\newcommand{\Qx}{\textsf{QX}}
\newcommand{\Qy}{\textsf{QY}}
\newcommand{\Qz}{\textsf{QZ}}
\newcommand{\QSNB}{\textsf{Q10}}
\newcommand{\sjoin}{\textsf{SJoin}}
\newcommand{\sjoinfk}{\textsf{SJoin\textunderscore opt}}
\newcommand{\cut}[1]{{}}
\newcommand{\xiao}[1]{{\color{black}{#1}}}
\newcommand{\binyang}[1]{{\color{black}{#1}}}
\usetikzlibrary{topaths,calc}
\tikzset{fontscale/.style = {font=Large}}
    
\lstdefinestyle{mystyle}{
	commentstyle=\color{green},
	keywordstyle=\color{purple},
	stringstyle=\color{purple},
	basicstyle=\small\ttfamily,
	breaklines=true,
	columns=fullflexible,
	frame=single
}

\begin{document}
\title{Reservoir Sampling over Joins}

\begin{abstract}
Sampling over joins is a fundamental task in large-scale data analytics. Instead of computing the full join results, which could be massive, a uniform sample of the join results would suffice for many purposes, such as answering analytical queries or training machine learning models.  In this paper, we study the problem of how to maintain a random sample over joins while the tuples are streaming in.  Without the join, this problem can be solved by some simple and classical reservoir sampling algorithms.  However, the join operator makes the problem significantly harder, as the join size can be polynomially larger than the input. 
We present a new algorithm for this problem that achieves a near-linear complexity. The key technical components are a generalized reservoir sampling algorithm that supports a predicate, and a dynamic index for sampling over joins.
We also conduct extensive experiments on both graph and relational data over various  join queries, and the experimental results demonstrate significant performance improvement  over the state of the art.
\end{abstract}

\author{Binyang Dai}
\authornote{Both authors contributed equally to this research.}
\email{bdaiab@connect.ust.hk}
    \orcid{0009-0000-5438-7288}
    \affiliation{%
    \institution{Hong Kong University of Science and Technology}
    \streetaddress{Clear Water Bay}
    \city{Hong Kong}
    \country{China}
    }

\author{Xiao Hu}
\authornotemark[1]
    \email{xiaohu@uwaterloo.ca}
    \orcid{0000-0002-7890-665X}
    \affiliation{%
    \institution{University of Waterloo}
    \streetaddress{200 University Ave W}
    \city{Waterloo}
    \state{Ontario}
    \country{Canada}
    \postcode{N2L 3G1}
    }

\author{Ke Yi}
\email{yike@ust.hk}
    \orcid{0000-0002-2178-3716}
    \affiliation{%
    \institution{Hong Kong University of Science and Technology}
    \streetaddress{Clear Water Bay}
    \city{Hong Kong}
    \country{China}
    }

\renewcommand{\shortauthors}{Bingyang Dai, Xiao Hu, and Ke Yi}



\setcopyright{acmlicensed}
\acmJournal{PACMMOD}
\acmYear{2024} \acmVolume{2} \acmNumber{3 (SIGMOD)} \acmArticle{118} \acmMonth{6}\acmDOI{10.1145/3654921}

\maketitle
\thispagestyle{plain}
\pagestyle{plain}

\section{Introduction}
\label{sec:intro}

In large-scale data analytics, people often need to compute complicated functions on top of the query results over the underlying relational database. 
However, the join operator presents a major challenge, since the join size can be polynomially larger than the original database. Computing and storing the join results is very costly, especially as the data size keeps increasing. Sampling the join results is thus a common approach used in many complicated analytical tasks while providing provable statistical guarantees.  One naive method is to first materialize the join results in a table and then randomly access the table, but this loses the performance benefit of sampling.  In as early as 1999, two prominent papers ~\cite{chaudhuri1999random, acharya1999join} asked the intriguing question, whether a sample can be obtained without computing the full join. As observed in by~\cite{chaudhuri1999random}, the main barrier is that the sampling operator cannot be pushed down, i.e., $\textsf{sample}(R \Join S) \neq \textsf{sample}(R) \Join \textsf{sample}(S)$. To overcome this barrier, the idea is to design some index structures to guide the sampling process.  Notably, an index was proposed for \textit{acyclic} joins (formally defined in Section~\ref{sec:acyclic}) that can be built in $O(N)$ time, where $N$ is the number of tuples in the database, which can then be used to draw a sample of the join results in $O(1)$ time~\cite{zhao2018random,carmeli2020answering}. Please see Section~\ref{sec:previous} for a more comprehensive review on the sampling complexity for different join queries.

This problem becomes more challenging in the streaming setting, where input tuples arrive at a high velocity.
How to efficiently and continuously maintain a uniform sample of the join results produced by tuples seen so far? One naive solution would be to re-build the index and re-draw the samples after each tuple has arrived, but this results in a total running time of $O(N^2)$ to process a stream with $N$ tuples.    
Recently, \citet{zhao2020efficient} applied the reservoir sampling algorithms~\cite{vitter1985random, li1994reservoir} to this problem to update the sample incrementally.  However, their index also suffers from a high maintenance cost that still leads to a total running time of $O(N^2)$ in the worst case.

This paper presents a new reservoir sampling algorithm for maintaining a sample over joins with a near-linear running time of $O(N\log N + k\log N \log{N \over k}),$ where $k$ is the given sample size.  Our algorithm does not need the knowledge of $N$; equivalently speaking, it works over an unbounded stream, and the total running time over the first $N$ tuples, for every $N \in \mathbb{Z}^+$, satisfies the  aforementioned bound.  This result is built upon the following two key technical ingredients, both of which are of independent interest. 

\paragraph{Reservoir sampling with a predicate}
The classical reservoir sampling algorithm, attributed to Waterman by Knuth \cite{knuth1997art}, maintains a sample of size $k$ in $O(N)$ time over a stream of $N$ items, which is already optimal.  Assuming there is a \sskip($i$) operation that can skip the next $i$ items \xiao{and jump directly to the next $(i+1)$-th item} in $O(1)$ time, the complexity can be further reduced to $O(k\log{N\over k})$, and there are several algorithms achieving this \cite{vitter1985random,li1994reservoir}. In this paper, we design a more general reservoir sampling algorithm that, for a given predicate $\theta$, maintains a sample of size $k$ only over the items on which $\theta$ evaluates to true.  The complexity of our algorithm is \binyang{$O\left(\sum_{i=1}^N\min\big(1, {k\over r_i + 1}\big)\right)$}, where $r_i$ the number of items in the first \binyang{$i-1$} items that pass the predicate.  Note that when $\theta(\cdot) \equiv \mathtt{true}$, we have \binyang{$r_i=i-1$} and the bound simplifies to $O(k \log {N\over k})$, matching the classical result.  Meanwhile, the complexity degrades gracefully as the stream becomes sparser, i.e., less items pass the predicate.  Intuitively, sparse streams are more difficult, as it is not safe to skip items.  In the extreme case where only one item passes the predicate, then the algorithm is required to return that item as the sample, and we have to check every item in order not to miss it.  

However, the assumption that \sskip($i$) takes $O(1)$ time is usually not true: One has to at least use a counter to count how many items have been skipped, \xiao{which already takes $O(i)$ time}. Interestingly, the reservoir sampling over joins problem provides a nice scenario where this assumption \textit{is} true, except that it has an $O(\log N)$ cost.  It is known  \cite{AGM2008} that the join results of $N$ tuples can be as many as $N^{\rho^*}$, where $\rho^*$ is the fractional edge cover number of the join (details provided in Section \ref{sec:cyclic}).  Thus, the stream of input tuples implicitly defines a polynomially longer \xiao{(conceptual)} stream of join results, which we want to sample from. As there is good structure in the latter, \xiao{there is no need to materialize this simulated join result stream; and moreover} it is possible to skip its items without counting them one by one. 

\paragraph{Dynamic sampling from joins}
Let $\Q$ be an acyclic join query, $\R$ a database instance of size $N$, and $\Q(\R)$ the join results of $\Q$ on $\R$. The second technical ingredient is an index structure that supports the following operations:
\begin{enumerate}[leftmargin=*]
    \item After a tuple is added to $\R$, the index structure can be updated in $O(\log N)$ time amortized.
    \item The index implicitly defines an array $J$ that contains $\Q(\R)$ plus some dummy tuples, but it is guaranteed that $|J| = O(|\Q(\R)|)$, i.e., the dummy tuples are no more than a constant fraction.  For any given $j\in[|J|]$, the index can return $J[j]$ in $O(\log N)$ time.  It can also return $|J|$ in $O(1)$ time. 
    \item The above is also supported for the delta query $\Delta \Q(\R, t):= \Q(\R \cup \{t\}) - \Q(\R)$ for any tuple $t \not\in \R$. 
\end{enumerate}
Note that operation (2) above directly solves the join sampling problem: We simply generate a random $j \in [|J|]$ and find $J[j]$, and repeat if it is dummy.  Since $|J| = O(|\Q(\R)|)$, this process will terminate after $O(1)$ trials in expectation, so the time to draw a sample is $O(\log N)$ expected.  This is only slightly slower than the previous index structures \cite{zhao2018random,carmeli2020answering}, which are inherently static.  Furthermore, operations (1) and (2) together also provide a solution for the reservoir sampling over join problem: For each tuple, we first update the index in $O(\log N)$ time and then re-draw $k$ samples in $O(k \log N)$ time.  This leads to a total time of $O(Nk\log N)$, already better than \cite{zhao2020efficient}, but still not near-linear.

To achieve near-linear time, we use operation (3) in conjunction with our reservoir sampling algorithm.  The observation is that each incoming tuple $t$ adds a batch of join results, which are defined by the delta query $\Delta \Q(\R,t)$.  If we can access any tuple in $\Delta \Q(\R,t)$ by position, then we can  implement a \sskip~ easily.  Our index can almost provide this functionality, except that it does so over $\Delta J$, which is a superset of $\Delta \Q(\R,t)$ that contains some dummy tuples.  This is exactly the reason why we need a reservoir sampling algorithm that supports a predicate.  We will run it over the stream of batches, where each batch is the $\Delta J$ of the corresponding delta query.  The predicate evaluates to true for the real tuples while false for the dummies. Finally, since each batch is dense (at least a constant \xiao{fraction} is real), our reservoir sampling algorithm will have good performance. 

\medskip
The contributions of this paper are thus summarized as follows:
\begin{itemize}[leftmargin=*]
    \item {\bf (Section~\ref{sec:reservoir})} We formulate the problem of reservoir sampling with a predicate.  Assuming \sskip~takes $O(1)$ time, we design an algorithm that can maintain a sample (without replacement) of size $k$ in time \binyang{$O\left(\sum_{i=1}^N\min\big(1, {k\over r_i + 1}\big)\right)$}, which we also show is instance-optimal. 
    \item {\bf (Section~\ref{sec:acyclic})} We present a dynamic index for acyclic joins that can be updated in $O(\log N)$ amortized time, and return a sample from either the full query or the delta query in $O(\log N)$ time.  Combined with our reservoir sampling with predicate algorithm, we show how the reservoir sampling over join problem can be solved in time $O(N \log N + k \log N \log \frac{N}{k})$. \binyang{We show how our algorithm can be optimized when key constraints are present. } 
    \item {\bf (Section~\ref{sec:cyclic})} We extend the algorithm to cyclic joins using the generalized hypertree decomposition technique.  In this case, the running time becomes $O(N^\textsf{w} \log N + k \log N \log \frac{N}{k})$, where $\textsf{w}$ is the fractional hypertree width of the query.
    \item {\bf (Section~\ref{sec:experiment})}  We implement our algorithm and evaluate it over both graph and relational data. The experimental results show that our algorithm significantly outperforms the state-of-the-art solution \cite{zhao2020efficient}.
\end{itemize}

\section{Preliminaries}

\subsection{Problem Definition}
\label{sec:probdef}
We first recap some standard notation in relational algebra \cite{abiteboul1995foundations}. 
 A multi-way (natural) join query can be defined as a hypergraph $\Q = (\V,\E)$, where $\V$ is the set of attributes, and $\E \subseteq 2^{\V}$ is the set of  relations. Let $\dom(x)$ be the domain of attribute $x \in \V$. A database instance $\R$ consists of a relation instance $R_e$ for each $e \in \E$, which is a set of tuples and each tuple $t \in R_e$ specifies a value in $\dom(v)$ for each attribute $v \in e$. For a tuple $t$, we use $\textsf{supp}(t)$ to denote the support of $t$, i.e., the set of attributes on which $t$ is defined. For attribute(s) $x$ and tuple $t$ with $x \subseteq \textsf{supp}(t)$, the projection $\pi_x t$ is the value of tuple $t$ on attribute(s) $x$.  
 The join results of $\Q$ over instance $\mathcal{R}$, denoted by $\Q(\R)$, is the set of all combinations of tuples, one from each $R_e$, that share common values for their common attributes, i.e., 
\begin{equation}
\label{eq:QR}
    \Q(\R) = \left\{t \in \prod_{x \in \V} \dom(x) \mid \forall e \in \E, \exists t_e \in R_e, \pi_{e} t = t_e\right\}.
\end{equation}  

For relation $R_e$ and tuple $t$, the semi-join $R_e \ltimes t$ returns the set of tuples from $R_e$ which have the same value(s) on attribute(s) $e \cap \textsf{supp}(t)$ with $t$. For a pair of relations $R_e, R_{e'}$, the semi-join $R_e \ltimes R_{e'}$ is the set of tuples from $R_e$ which has the same value(s) on attribute(s) $e \cap e'$ with at least one tuple from $R_{e'}$.  Note that for a join query $\Q$, the delta query $\Delta \Q(\R,t)$ is equal to $\Q(\R \cup \{t\}) \ltimes t$.

In the streaming setting, we model each tuple as a triple $u = (t, i, R_e)$ for $i \in \mathbb{Z}^+$, indicating that tuple $t$ is inserted into relation $R_e$ at time $i$. Let $D$ be the stream of input tuples, ordered by their timestamp. 
Let $\R^i$ be the database defined by the first $i$ tuples of the stream, and set $\R^0 = \emptyset$.  We use $N$ to denote the length of the stream, which is only used in the analysis.  The algorithms will not need the knowledge of $N$, so they work over an unbounded stream. 

There are two versions of the join sampling problem: The first is the sampling over join problem as studied in \cite{acharya1999join, chaudhuri1999random, zhao2018random, carmeli2020answering, chen2020random, deng2023join, Kim2023guaranteeing}.  This is an indexing (data structure) problem where we wish to have an index that supports drawing a sample from $\Q(\R^i)$.  For this problem, we care about the sampling time $t_s$ and the update time $t_u$. For a static index, we care about the index construction time and the sampling time.  The other is the reservoir sampling problem, as studied in \cite{vitter1985random,li1994reservoir,zhao2020efficient}, where we wish to maintain $k$ random samples from $\Q(\R^i)$ without replacement for every $\Q(\R^i), i\in \mathbb{Z}^+$.  For this problem, we just care about the total running time.  Note that any solution for the former yields a solution for the latter with total time $O(t_u \cdot N + t_s \cdot Nk)$, but this may not be optimal.  For both versions of the problem, all algorithms, including ours, use $O(N)$ space.  Note that the classical reservoir sampling only uses $O(k)$ space, but sub-linear space is not possible when $\Q$ has  joins. Just consider a two-table join $\Q:=R_1(X,Y) \Join R_2(Y,Z)$.  Suppose the first $N$ tuples in the stream are all in $R_1$.  The algorithm must keep all of them in memory, otherwise it will miss the first join result, which must be sampled, when some tuple in $R_2$ arrives. 

We follow the convention of data complexity~\cite{abiteboul1995foundations} and analyze the running time in terms of the input size $N$ and sample size $k$, while taking the size of $\Q$ (i.e., $|\V|$ and $|\E|$) as a constant.  
We follow the set semantics, so inserting a tuple into a relation that already has it has no effect.  Thus we assume that duplicates have been removed from the input stream.

\subsection{Previous Results}
\label{sec:previous}
    \noindent {\bf Sampling over joins.} Chaudhuri et al. \cite{chaudhuri1999random} showed, for the basic two-table join $R_1(x_1,x_2) \Join R_2(x_2,x_3)$, how to construct an index structure in $O(N)$ time, such that a sample can be drawn in $O(1)$ time.  Acharya et al.~\cite{acharya1999join} achieved the same complexity result for multi-way joins but all joins are restricted to foreign-key joins.  These results have been later extended to all acyclic joins \cite{zhao2018random,carmeli2020answering}. 
    All these sampling indexes on acyclic joins are static.
    For cyclic joins, there is an index that can be built in $O(N)$ time, while a sample can be drawn in $O\Big(\frac{N^{\rho^*+1}}{|\Q(\R)|}\Big)$ time \cite{chen2020random}, which has recently been improved to  $O\Big(\frac{N^{\rho^*}}{|\Q(\R)|}\Big)$~\cite{deng2023join, Kim2023guaranteeing}, who also make the index dynamic, but note that the sampling time for cyclic joins is significantly higher than that for acyclic joins. 

    \paragraph{Reservoir sampling over joins} When $\Q$ has no joins, the classic reservoir sampling algorithm~\cite{knuth1997art} solves the problem in $O(N)$ time.  Assuming that \sskip~takes $O(1)$ time, this can be reduced to $O(k \log {N \over k})$~\cite{vitter1985random, li1994reservoir}. 
    Zhao et al.~\cite{zhao2020efficient} investigated the problem over acyclic joins, and proposed some efficient heuristics.  But their solution takes $O(N^2)$ time in the worst case, which is the same as the naive solution that rebuilds the static join sampling index \cite{zhao2018random,carmeli2020answering} at each time step. 
     
   \paragraph{Hardness results}
    In this paper we will focus on acyclic joins over an insertion-only stream.  Both restrictions turn out to be necessary for achieving a near-linear running time, following some existing hardness results.  The observation is that sampling a query is at least as hard as the corresponding Boolean query (i.e., determining whether $\Q(\R)=\emptyset$): We can just return true for the Boolean query if there is any sample returned from the sampling algorithm, and false otherwise.  It is known that it requires $\Omega(N^{\mathsf{w}})$ time to compute a Boolean cyclic query, for some width parameter $\mathsf{w}>1$ of the query~\cite{bagan2007acyclic}. So for a cyclic query, there is no hope for taking a sample of the join in near-linear time, even over a static database.  Meanwhile, for a general (more precisely, a non-hierarchical) acyclic query, it is known that the update time must be $\Omega(\sqrt{N})$ just to maintain the Boolean answer, when both insertions and deletions are allowed~\cite{berkholz17:_answer}.  This means that the reservoir sampling problem requires at least $\Omega(N^{1.5})$ time over a fully-dynamic stream. 

\section{Reservoir Sampling with Predicate}
\label{sec:reservoir}

\subsection{Reservoir Sampling Revisited}
\label{sec:reservoir-revisit}
Reservoir sampling~\cite{vitter1985random, li1994reservoir} is a family of algorithms for maintaining a random sample, without replacement, of $k$ items from a possibly infinite stream. 
The classical version, as described in \cite{knuth1997art}, works as follows. {\bf (Step 1)} It initializes an array $S$ (called the {\em reservoir}) of size $k$, which contains the first $k$ items of the input. {\bf (Step 2)} For each new input $x_i$, it generates a random number $j$ uniformly in $[1,i]$. If \binyang{$j \le k$}, then it replaces $S[j]$ with $x_i$. Otherwise, it simply discards $x_i$. At any time, $S$ is a uniform sample without replacement of $k$ items of all items processed so far. Clearly, this algorithm takes $O(N)$ time to process a stream of $N$ items.  Also, the algorithm does not need the knowledge of $N$, so it works over an unbounded stream.

Assuming a \sskip($i$) operation that can skip the next $i$ items in $O(1)$ time, more efficient versions are known.  In particular, we will make use of the one from \cite{li1994reservoir}.  It is based on the fact that, in a set of $N$ independent random numbers drawn the uniform distribution $\textsf{Uni}(0,1)$, the indices of the smallest $k$ random numbers are a sample without replacement from the index set $\{1,2,\cdots,N\}$. The algorithm works as follows. {\bf (Step 1)} It initializes $S$ as before, and set $w= u^{1/k}$ for $u \sim \textsf{Uni}(0,1)$.  {\bf (Step 2)} It draws a random number $q$ from the geometric distribution $\textsf{Geo}(w)$, and skip the next $q$ items. It then replaces a random item from $S$ with $x_i$, and updates $w$ to $w \cdot u^{1/k}$ for $u \sim \textsf{Uni}(0,1)$. It can be shown \cite{li1994reservoir} that at any time, $S$ is a sample without replacement of $k$ items of all items processed so far, and this algorithm runs in $O(k \cdot \log \frac{N}{k})$ expected time, which is optimal.
\begin{algorithm}[t]   
\caption{\textsc{Reservoir}$(D,k,\theta)$}
    \label{alg:reservoir}

    \SetKwInOut{Input}{Input}
    \SetKwInOut{Output}{Output}

    \Input{An input stream $D$ of items, an integer $k >0$, and a predicate $\theta$;}
    \Output{A set $S$ maintaining $k$ random samples without replacement of items on which $\theta$ evaluates to true;}
    
    $S \gets \emptyset$\;
    \While{$|S| < k$}{
        $x \gets D.\nnext()$\;
        \lIf{$x = \nnull$}{\textbf{break}} 
        \lIf{$\theta(x)$}{
        $S \gets S \cup \{x\}$}
    }
    $w \gets \random()^{1/k}$\;
    $q \gets \lfloor(\ln(\random()/\ln(1-w))\rfloor$\;
    \While{$\textup{\textbf{true}}$}{
        $x \gets D.\sskip(q)$\;
        \lIf{$x = \nnull$}{\textbf{break}}
        \If{$\theta(x)$}{
            $y \gets$ a randomly chosen item from $S$\;
            $S \gets S - \{y\} + \{x\}$\;
            $w \gets w \cdot \random()^{1/k}$\;
        }
        $q \gets \lfloor(\ln(\random()/\ln(1-w))\rfloor$; (note that $q\sim \textsf{Geo}(w)$)\ 
    }
\end{algorithm}

\subsection{Reservoir Sampling with Predicate}

The problem of {\em reservoir sampling with predicate} is defined as follows. Given an input stream of items, a predicate $\theta$ and an integer $k > 0$, it asks to maintain a sample of size $k$ of all items on which $\theta$ evaluates to true (these items are also called real items, while the others dummy).  We assume that $\theta$ can be evaluated in $O(1)$ time.   Note that the $O(N)$-time algorithm easily supports a predicate: We just evaluate $\theta$ on each item and feed the real items to the algorithm.  It is more nontrivial to adapt the $O(k\log{N \over k})$ algorithm, since the \sskip~operation skips an unknown number of  real items.

We adapt the reservoir sampling algorithm~\cite{li1994reservoir} to Algorithm~\ref{alg:reservoir}.  In the description, we use the following two primitives:
\begin{itemize}[leftmargin=*]
    \item {\bf \nnext()} returns the next item if it exists, and $\nnull$ otherwise;
    \item {\bf \sskip($i$)} skips the next $i$ items and returns the $(i+1)$-th item if it exists, and $\nnull$ otherwise.
\end{itemize}

Compared with \cite{li1994reservoir}, we have made two changes: (line 2-5) when the reservoir is not full, we only add real items to it; (line 11 - 14) we only update the reservoir and the parameter $w$ when the algorithm stops at a real item. The correctness proof of Algorithm~\ref{alg:reservoir} given below is quite technical. We first provide some intuition here. For each item $x$ in the stream, Algorithm~\ref{alg:reservoir} generates a random variable from $\textsf{Uni}(0,1)$, say $u_x$, and executes line 11-14 for $x$ if $u_x < w$, since the next random variable generated from $\textsf{Uni}(0,1)$ being smaller than $w$ follows the geometric distribution parameterized by $w$. We can further exchange the two if-conditions ($w < u$ and $x$ is real) since these two events are independent. Then, one can show that Algorithm~\ref{alg:reservoir} is equivalent to feeding only the real items to the non-predicate version of the algorithm in \cite{li1994reservoir}.

\begin{theorem}
    \label{the:reservoir-correctness}
    Algorithms~\ref{alg:reservoir} correctly maintains a uniform sample of $k$ real items.
\end{theorem}

\begin{proof}
Algorithm~\ref{alg:reservoir} runs essentially the same process as Algorithm~\ref{alg:naive-reservoir-I}, since the next random value drawn from $\textsf{Uni}(0,1)$ being smaller than $w$ follows the geometric distribution parameterized by $w$. 
\begin{algorithm}[h]    
\caption{\textsc{NaiveReservoir-I}$(D,k)$}
    \label{alg:naive-reservoir-I}

    \SetKwInOut{Input}{Input}
    \SetKwInOut{Output}{Output}

    \Input{An input stream $D$ of elements, an integer $k>0$ and a predicate $\theta$;}
    \Output{A set $S$ maintaining $k$ uniform samples without replacement of items that pass the predicate $\theta$;}
    
    $S \gets \emptyset$\;
    \While{$|S| < k$}{
        $x \gets D.\nnext()$\;
        \lIf{$x = \nnull$}{\textbf{break}}
        \lIf{$\theta(x)$}{$S \gets S \cup \{x\}$}
    }
     $w \gets \random()^{1/k}$\;
    \While{$\textup{\textbf{true}}$}{
        $x \gets D.\nnext()$\;
        \lIf{$x = \nnull$}{\textbf{break}}
        $u_x \gets \random()$\;
        \If{$u_x < w$}{
            \If{$\theta(x)$}{
                $y \gets$ a randomly chosen element from $S$\;
                $S \gets S - \{y\} + \{x\}$\;
                $w \gets w \cdot \random()^{1/k}$\;
            }
        }
    }
\end{algorithm}
\begin{algorithm}[t]
\caption{\textsc{NaiveReservoir-II}$(D,k)$}
    \label{alg:naive-reservoir-II}

    \SetKwInOut{Input}{Input}
    \SetKwInOut{Output}{Output}

    \Input{An input stream $D$ of elements, an integer $k>0$ and a predicate $\theta$;}
    \Output{A set $S$ maintaining $k$ uniform samples without replacement of items that pass the predicate $\theta$;}
    
    $S \gets \emptyset$\;
    \While{$|S| < k$}{
        $x \gets D.\nnext()$\;
        \lIf{$x = \nnull$}{\textbf{break}}
        \lIf{$\theta(x)$}{$S \gets S \cup \{x\}$}
    }
     $w \gets \random()^{1/k}$\;
     \While{$\textup{\textbf{true}}$}{
        $x \gets D.\nnext()$\;
        \lIf{$x = \nnull$}{\textbf{break}}
        $u_x \gets \random()$\;
        \If{$\theta(x)$}{
            \If{$u_x < w$}{
                $y \gets$ a randomly chosen element from $S$\;
                $S \gets S - \{y\} + \{x\}$\;
                $w \gets w \cdot \random()^{1/k}$\;
            }
        }
    }
\end{algorithm}
Furthermore, in Algorithm~\ref{alg:naive-reservoir-I}, it is safe to exchange the if condition in line 11 with the if condition in line 12, since these two conditions are independent of each other. After this exchange, we obtain Algorithm~\ref{alg:naive-reservoir-II}. Note that Algorithm~\ref{alg:naive-reservoir-II} simply runs the classic reservoir sampling only over real elements, whose correctness is proved in~\cite{li1994reservoir}. Together, the correctness of Algorithm~\ref{alg:reservoir} is proved by the correctness of the classic reservoir sampling algorithm.
\end{proof}

The time complexity of Algorithm~\ref{alg:reservoir} depends on how the real and dummy items are distributed in the stream, as more precisely characterized by the following theorem:

\begin{theorem}
\label{the:reservoir}
    Algorithm~\ref{alg:reservoir} runs in \binyang{$O(\alpha \cdot (p-1) + 
    \gamma \cdot \sum_{i=p}^N \frac{k}{r_i + 1})$} expected time over a stream of $N$ items, where \binyang{$r_i$ is the number of real items in the first $i-1$ items, $p$ is the smallest $i$ such that $r_i= k$ (set $p=N+1$ if no such $p$ exists)}, and $\alpha$ and $\gamma$ are the time complexities of $\nnext(\cdot)$ and $\sskip(\cdot)$, respectively.
\end{theorem}

\begin{proof}
    \binyang{For the ease of analysis, we add one additional real item as a sentinel object to the end of the stream. We will ignore the cost for this real item later.} Let $S'$ denote the set of items with index \binyang{larger than or equal to} $p$. We analyze the number of invocations of $\sskip(\cdot)$ by Algorithm~\ref{alg:reservoir}. Each time Algorithm~\ref{alg:reservoir} calls $\sskip(\cdot)$ and returns some item $x$, we say it stops at $x$. \xiao{Note that Algorithm~\ref{alg:reservoir} only stops at items that are from $S'$.} We consider an equivalent version for the ease of analysis: for each item, we generate a random variable from $\textsf{Uni}(0,1)$, say $u$, and execute line 11 - 14 only for successful trials (i.e. $u < w$). The number of stops is the same as the number of successful trials. If $w_i$ is the value of $w$ when processing item $x_i$, then the probability that Algorithm~\ref{alg:reservoir} stops at $x_i$ is exactly $w_i$ for \binyang{$i \ge p$}. Let \binyang{$\mathbf{w} = \langle w_1,w_2,\cdots, w_{N+1} \rangle$} be the state of Algorithm~\ref{alg:reservoir}. For any $w$, we observe that (1) $w_i \in [0,1]$ if \binyang{$i \in [1\dots N+1]$}; (2) $w_i \ge w_j$ if $i < j$; (3) $w_i =1$ if \binyang{$i < p$}; (4) \binyang{$w_i = w_j$ if the $j$-th item is real and the $i$-th item to the $(j-1)$-th item are all dummy. For (4), we use function $\pi(i)$ to denote the smallest index $j$ where $j \ge i$ such that $x_j$ is real. Note that $w_i = w_{\pi(i)}$ for $i \in [p\dots N+1]$.} Let $W$ be the set of all possible states of Algorithm~\ref{alg:reservoir}. The expected number of stops in $S'$ is 
    \begin{align*}
        \mathbb{E}[\# \textsf{stops in } S']  & = \sum_{\mathbf{w} \in W}\Pr(\mathbf{w}) \cdot \mathbb{E}[\# \textsf{stops in } S'|\mathbf{w}] \\
        & = \sum_{\mathbf{w} \in W} \Pr(\mathbf{w}) \cdot \binyang{\sum_{i = p}^N} w_{\pi(i)} = \binyang{\sum_{i = p}^N} \sum_{\mathbf{w} \in W}\Pr(\mathbf{w}) \cdot w_{\pi(i)}  
        = \binyang{\sum_{i = p}^N \frac{k}{r_i + 1}}
    \end{align*} 
    where the rationale behind the last equality is that $\sum_{\mathbf{w} \in W}\Pr(\mathbf{w}) \cdot w_{\pi(i)}$ is exactly the probability that $x_{\pi(i)}$ enters the reservoir ever.  Next, we move to the $\nnext()$ primitive. For while loop in line 2, we keep adding real items into the reservoir until it becomes full. So, the number of invocations of the $\nnext()$ is exactly \binyang{$p-1$}. Putting everything together, we complete the proof. 
\end{proof}

Note that in the degenerate case where all items are real, we have \binyang{$r_i=i-1$} and the running time of Algorithm \ref{alg:reservoir} becomes $O(k \log{N \over k})$ (when taking $\alpha, \gamma$ as $O(1)$), matching the optimal reservoir sampling running time \cite{vitter1985random,li1994reservoir}.  In the other extreme case, all items are dummy, so \binyang{$p=N+1$} and $r_i=0$, and the running time becomes $O(N)$, i.e., no item is skipped.  Indeed in this case, it is not safe to skip anything; otherwise, the algorithm may miss the first real item if one shows up, which must be sampled. 
Below, we formalize this intuition and prove that Algorithm \ref{alg:reservoir} is not just optimal in these two degenerate cases, but in all cases, namely, it is instance-optimal. 

\begin{theorem}
\label{the:optimality}
    For any input stream $S$ of \binyang{$N$} elements, any algorithm that can maintain a uniform sample of size $k$ over all real elements must run in \binyang{$\Omega\left(\sum_{i=1}^N \min\{1, \frac{k}{r_i+1}\}\right)$} expected time.
\end{theorem}

\begin{proof}
    Consider an arbitrary input stream $S$ and an arbitrary $i \in [N]$. 
    Any correct algorithm for maintaining a uniform sample over real elements at timestamp $i$ must stop at $x_i$ with probability \binyang{at least $\min\{1, \frac{k}{r_i+1}\}$}. Recall that any algorithm cannot distinguish whether an element is real or dummy until it stops (and checks). Suppose $x_i$ is real. If the probability is smaller than \binyang{$\min\{1, \frac{k}{r_i + 1}\}$}, then the probability that $x_i$ enters into the reservoir must be smaller than \binyang{$\min\{1, \frac{k}{r_i + 1}\}$}, which contradicts the fact that this algorithm can return a uniform sample at timestamp $i$. A correct algorithm for maintaining a uniform sample of size $k$ over all real elements  in the stream must be correct at timestamp $i$ for $1\le i \le N$. Hence, any correct algorithm must stops at \binyang{$\sum_{i=1}^N \min\{1, \frac{k}{r_i + 1}\}$} expected numbers.  
\end{proof}

Although the running time of Algorithm \ref{alg:reservoir} can vary significantly from $O(k \log{N \over k})$ to $O(N)$, but it is closer to the former as long as the stream is dense enough.

\begin{definition}[Dense stream]
\label{def:dense_seq}
    Given a stream $S = \langle x_1, x_2, \cdots, x_n \rangle$, $S$ is \textit{$\phi$-dense} for $0< \phi \le 1$, if \binyang{$r_i \ge \phi \cdot (i-1)$} for all $i$.
\end{definition}

Combining Theorem~\ref{the:reservoir} and Definition~\ref{def:dense_seq} we obtain: 

\begin{corollary}
\label{cor:reservoir-dense}
For any $\phi$-dense stream where $\phi$ is a constant,  Algorithm~\ref{alg:batch-reservoir} runs in $O(\alpha \cdot k + \gamma \cdot k\log \frac{N}{k})$ expected time.
\end{corollary}

We also mention three important properties for dense streams, which will be used later for joins. \xiao{Lemma~\ref{lem:dense_concat} implies that straightforwardly concatenating two streams still preserves their minimum density of real items. If one stream only consists of dummy items, it is possible to get a better bound on the density of real items in the whole stream, which is essentially captured by Lemma~\ref{lem:dense_padding}. The more dummy items padded, the sparser the stream becomes. Lemma~\ref{lem:dense_cross_product} implies that mixing two streams as their Cartesian product preserves a density that is at least half of their density product. } \binyang{For the ease of notation, we denote $q_i = r_{i+1}$ (i.e., the number of real items in the first $i$ items) in the following proofs of Lemma~\ref{lem:dense_concat}, Lemma~\ref{lem:dense_cross_product}, and Lemma~\ref{lem:dense_padding}. It is easy to see that a stream $S$ is $\phi$-dense if $q_i \ge \phi \cdot i$ for all $i$.}

\begin{lemma}
\label{lem:dense_concat}
    Given two streams $S_1 = \langle x_{1}, x_{2}, \cdots x_{m} \rangle$ and $S_2 = \langle y_{1}, y_{2}, \cdots, y_{n} \rangle$, if $S_1$ is $\phi_1$-dense and $S_2$ is $\phi_2$-dense, their concatenation $S_1 \circ S_2 := \langle x_{1}, x_{2}, \cdots, x_{m}, y_{1}, y_{2}, \cdots, y_{n}\rangle$ is $\min\{\phi_1,\phi_2\}$-dense.
\end{lemma}

\begin{proof}
Consider the stream $S_1 \circ S_2$. As $S_1$ is $\phi_1$-dense, we have \binyang{$q_i \ge \phi_1 \cdot i$} for every $i \in [1\dots m]$. Moreover, \binyang{$q_m \ge \phi_1 \cdot m$}. As $S_2$ is $\phi_2$-dense, we have \binyang{$q_j - q_m \ge \phi_2 \cdot (j - m)$} for every $m \le j \le m+n$. Hence, we obtain \binyang{
\[q_j \ge \phi_2 \cdot (j - m) + q_m \ge \phi_2 \cdot (j - m) + \phi_1 \cdot m \ge \min\{\phi_1, \phi_2\} \cdot j\]}
for every $m \le j \le m+n$. Together with the fact \binyang{$q_j \ge \phi_1 \cdot j \ge  \min\{\phi_1, \phi_2\} \cdot j$} for every $j\in [1\dots m]$, we have proved that \binyang{$q_j \ge \min\{\phi_1, \phi_2\} \cdot j$} for every $j \in [1\dots m+n]$.
\end{proof}

\begin{lemma}
\label{lem:dense_cross_product}
    Given two streams $S_1 = \langle x_{1},x_{2},\dots, x_{m} \rangle$ and $S_2 = \langle y_{1},y_{2},\dots, y_{n} \rangle$, if $S_1$ is $\phi_1$-dense and $S_2$ is $\phi_2$-dense, their Cartesian product $S_1 \times S_2 := \langle (x_{1}, y_{1}), \cdots, (x_{1}, y_{n})$, $(x_{2}, y_{1}), \cdots, (x_{2}, y_{n})$, $(x_{m}, y_{1}), \cdots (x_{m}, y_{n})\rangle$ is $\left(\frac{\phi_1 \phi_2}{2}\right)$-dense, where $(x_i,x_j)$ is real if  and only if both $x_i$ and $x_j$ are real.
\end{lemma}

\begin{proof}
\binyang{Consider the stream $S_1 \times S_2 = \langle z_1,z_2, \dots z_{mn}\rangle$ and an arbitrary $1 \le i \le mn$. Assume that $m > 0$ and $n > 0$. Let $i_1 = \lfloor \frac{i}{n} \rfloor$ and $i_2 = i - i_1 \cdot n$. Let $\langle z_{jn+1},z_{jn+2},\dots z_{jn+n} \rangle$ be the $row_j$, where $j = 0,1,\dots i_1-1$. Then $row_j$ contains at least $\phi_2 \cdot n$ real items if $x_{j+1}$ in $S_1$ is real. Otherwise, all the $n$ items in $row_j$ are dummy. Let $q_i$ be the number of items that are real in the first $i$ items of the resulted stream. We distinguish the following 2 cases: 
\begin{itemize}[leftmargin=*]
    \item $i_1 = 0$: Since $S_1$ is $\phi_1$-dense, the first item of $S_1$ must pass predicate $\theta$. Then in this case, $q_i \ge \phi_2 \cdot i$ as $S_2$ is $\phi_2$-dense.
    \item $i_1 > 0$: The first $i$ items of the resulted stream can be represented as $i_1$ rows followed by $i_2$ items. In the $i_1$ rows, there are at least $\phi_1 \cdot i_1$ rows each contains at least $\phi_2 \cdot n$ items that are real as $S_1$ is $\phi_1$-dense. In total, there are at least $\phi_1 \cdot \phi_2 \cdot n \cdot i_1$ item that are real. As $i \le (i_1 + 1) \cdot n$, we have
    \begin{align*}
        q_i \ge \frac{\phi_1 \cdot \phi_2 \cdot n \cdot i_1}{i} \cdot i & \ge \frac{\phi_1 \cdot \phi_2 \cdot n \cdot i_1}{(i_1 + 1) \cdot n} \cdot i 
        \ge \phi_1 \cdot \phi_2 \cdot \frac{i_1}{i_1 + 1} \cdot i \ge \frac{\phi_1 \cdot \phi_2}{2} \cdot i
    \end{align*}
\end{itemize}  
As $0 < \phi_1 \le 1$, $q_i \ge \phi_2 \cdot i \ge \frac{\phi_1 \cdot \phi_2}{2} \cdot i$ for the case $i_1 = 0$. Putting all together, we have $q_i \ge \frac{\phi_1 \cdot \phi_2}{2} \cdot i$.}
\end{proof}

\begin{lemma}
\label{lem:dense_padding}
    \xiao{Given a $\phi$-dense stream of size $m$, padding $n$ dummy items at the end yields a $\left(\frac{m}{m+n} \cdot \phi\right)$-dense stream.}
\end{lemma}

\begin{proof}
    Let $S, S'$ be the original and resulted stream respectively. Let $|S| = m$ and $|S'| = m + n$, i.e., padding $n$ dummy items at the end of $S$. It suffices to prove $q_i \ge \phi \cdot \frac{m}{m+n} \cdot i$ for any $m+1 \le i \le m+n$. As $S$ is $\phi$-dense, we note that $q_m \ge \phi \cdot m$. We have
    \begin{align*}
        q_i &= q_m \ge \phi \cdot m \ge \phi \cdot \frac{m}{m+n} \cdot (m+n) \ge \phi \cdot \frac{m}{m+n} \cdot i \qedhere
    \end{align*} 
\end{proof}

\subsection{Batched Reservoir Sampling with Predicate}

As described in Section \ref{sec:intro}, each arriving tuple $t$ generates a batch of new join results $\Delta \Q(\R, t)$. 
To apply our reservoir sampling algorithm over joins, we first adapt Algorithm~\ref{alg:reservoir} into a batched version.  
Formally, given an input stream of item-disjoint batches $\langle B_1, B_2,\cdots, B_m\rangle$, and a predicate $\theta$, the goal is to maintain $k$ uniform samples without replacement from $B^{\theta}_1 \cup B^{\theta}_2 \cup \cdots \cup B^{\theta}_i$ for every $i$, where $B^{\theta}_i \subseteq B_i$ is the set of real items in batch $B_i$.  

The framework of our batched reservoir sampling is described in Algorithm~\ref{alg:batch-reservoir}, which calls $\textsc{BatchUpdate}$ (Algorithm~\ref{alg:batch-update}) for every batch. $\textsc{BatchUpdate}$ essentially runs Algorithm \ref{alg:reservoir} on the given batch $B$, but it must guard against the case where a \sskip($q$) may skip out of the batch.  For this purpose, it needs another primitive:
\begin{itemize}[leftmargin=*]
    \item {\bf \remain()} returns the number of remaining items in a batch.
\end{itemize}
More precisely, when $B.$\remain() $\le q$, we skip all the remaining items in the current batch, and pass $q-B.$\remain() as another parameter to the next batch so that the first $q-B.$\remain() items in the next batch will be skipped.  The details are given in Algorithm~\ref{alg:batch-update}.
\begin{algorithm}[t]
\caption{\textsc{BatchReservoir}$(D, k, \theta)$}
    \label{alg:batch-reservoir}

    \SetKwInOut{Input}{Input}
    \SetKwInOut{Output}{Output}

    \Input{An input stream $D$ of item-disjoint batches, an integer $k>0$, and a predicate $\theta$;}
    \Output{A set $S$ maintaining $k$ random samples without replacement of items on which $\theta$ evaluates to true;}
    
    $S \gets \emptyset$, \xiao{$w \gets +\infty$,} $q \gets 0$\;
    \ForEach{\textrm{batch} $B \in D$}{
        $(S,w,q) \gets \textsc{BatchUpdate}(S,k,B,q,w, \theta)$\;
    }
\end{algorithm}
\begin{algorithm}[t]
\caption{\textsc{BatchUpdate}$(S, k, B, q, w, \theta)$}
\label{alg:batch-update}
\SetKwInOut{Input}{Input}
\SetKwInOut{Output}{Output}
\Input{A set $S$ of random samples, an integer $k > 0$, a new batch $B$ with the first $q$ items to be skipped, \xiao{parameter $w$} and a predicate $\theta$;}
\Output{Updated $S$, $w$ and $q$;}
    \While{$|S| < k$ and $B.\remain() > 0$}{
        $x \gets B.\nnext()$\;
        \lIf{$\theta(x)$}{$S \gets S \cup \{x\}$}
    }
    \lIf{$|S|<k$}{\Return $S,w,q$}
    \xiao{
    \If{$w>1$}{
        $w \gets \random()^{1/k}$\;
        $q \gets \lfloor(\ln(\random())/\ln(1-w))\rfloor$; (note that $q\sim \textsf{Geo}(w)$)\
    }
    }
    \While{$B.\remain() > q$}{
        $x \gets B.\sskip(q)$\;
        \If{$\theta(x)$}{
            $y \gets $ a randomly chosen item from $S$\;
            $S \gets S - \{y\} + \{x\}$\;
            $w \gets w \cdot \random()^{1/k}$\;
        }
        $q \gets \lfloor(\ln(\random()/\ln(1-w))\rfloor$; (note that $q\sim \textsf{Geo}(w)$)\
    }
    \Return $S,w, q- B.\remain()$; \
\end{algorithm}
\begin{algorithm}[t]    
\caption{\textsc{ReservoirJoin}$(\Q, D, k)$}
\label{alg:reservoir-sampling-join}
    \SetKwInOut{Input}{Input}
    \SetKwInOut{Output}{Output}

    \Input{A join query $\Q$, an input stream $D$ of tuples, and the target number of samples $k$;}
    \Output{A set $S$ maintaining $k$ random samples without replacement for the join results of $\Q$ over tuples seen as far;}
    
    Initialize index $\mathcal{L}$, $S \gets \emptyset$, \xiao{$w \gets +\infty$,} $q \gets 0$, \binyang{$\theta \gets \textsf{isReal}$($\cdot$)}\;
    \While{\textup{\textbf{true}}}{
        $t \gets D.\nnext()$\;
        \lIf{$t = \nnull$}{\textbf{break}}
        $\mathcal{L} \gets \textsc{IndexUpdate}(\mathcal{L},t)$\;
        $B \gets \textsc{BatchGenerate}(\Q, \mathcal{L}, t)$\;
        \binyang{$(S,w,q) \gets \textsc{BatchUpdate}(S,k,w,q,B,\theta)$}\;
    }
\end{algorithm}
\xiao{Moreover, we note that parameters $w,q$ are only initialized once (as line 6-7 in Algorithm~\ref{alg:reservoir}), i.e., the first time when the reservoir $S$ is filled with $k$ items. To ensure this in the batched version, we set $w$ with $+\infty$ at the beginning (line 1 of Algorithm~\ref{alg:batch-reservoir}), so that $w,q$ will be initialized the first time when the reservoir $S$ is filled with $k$ items, and will never be initialized again no matter how many times Algorithm~\ref{alg:batch-update} is invoked, since the value of $w$ is always no larger than $1$ after initialization.} 

The samples maintained by Algorithm~\ref{alg:batch-reservoir} 
are exactly the same as that maintained by Algorithm~\ref{alg:reservoir} over items in batches, so correctness follows immediately.  Below we analyze its running time.

\begin{theorem}
\label{the:batch-reservoir}
    Given an input stream of batches each of which is $\phi$-dense for some constant $\phi$, Algorithm~\ref{alg:batch-reservoir} runs in $O((\alpha + \beta) \cdot k + (\beta + \gamma) \cdot k\log \frac{N}{k} + m)$ expected time over a stream of $m$ item-disjoint batches, where $N$ is the total size of items in all batches, and $\alpha, \beta, \gamma$ are the time complexities of $\nnext(\cdot)$, $\remain(\cdot)$, $\sskip(\cdot)$ respectively.
\end{theorem}

\begin{proof}
    Running Algorithm~\ref{alg:batch-reservoir} on a stream of batches containing $\phi$-dense streams is essentially the same as running Algorithm~\ref{alg:reservoir} on the concatenation of the $\phi$-dense streams. The numbers of invocation to $\nnext(\cdot)$ and $\sskip(\cdot)$ in Algorithm~\ref{alg:batch-reservoir} are the same as in Algorithm~\ref{alg:reservoir}. Note that each call to $\remain(\cdot)$ is immediately followed by a call to $\nnext(\cdot)$ or $\sskip(\cdot)$ except for line 15. In line 15, the return value of $\remain(\cdot)$ must be the same as the last call to $\remain(\cdot)$ in line 8. Hence, we can store it in a variable and eliminate the invocation in line 15. The last $O(m)$ term comes from the fact that Algorithm~\ref{alg:batch-reservoir} makes $m$ calls to \textsc{BatchUpdate} in total. Then, it follows Corollary~\ref{cor:reservoir-dense}. 
\end{proof}

\begin{figure*}[t]
    \centering
    \includegraphics[scale=0.72]{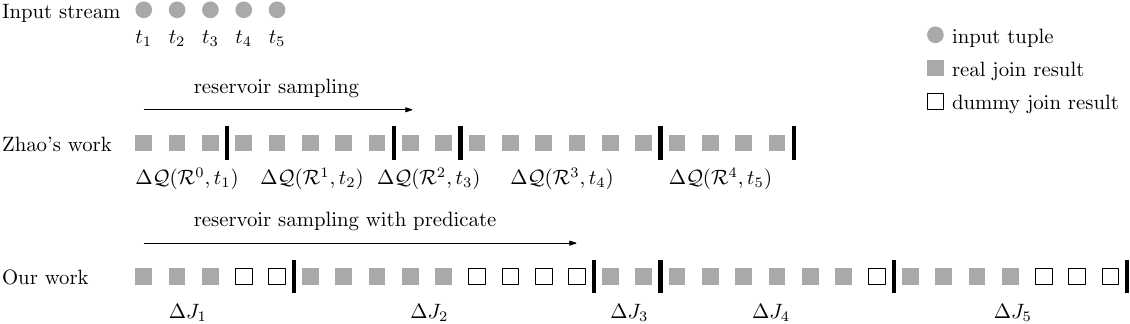}
    \vspace{-1em}
    \caption{An illustration of reservoir sampling over join.}
    \label{fig:framework}
\end{figure*}

\subsection{Reservoir Sampling Over Joins}
\label{sec:samplejoin}
The framework for reservoir sampling over joins is described in Algorithm~\ref{alg:reservoir-sampling-join}.
For each tuple $t$ in the input stream, we invoke procedure  $\textsc{BatchGenerate}$ to conceptually generate a batch $\Delta \Q(\R,t)$ and feed it into the batched reservoir sampling algorithm.  However, we cannot afford to materialize each batch, whose total size could be as large as $O(N^{\rho^*})$, where $\rho^*$ is the fractional edge cover number of join (details provided in Section~\ref{sec:cyclic}).  Instead, we will maintain a linear-size index $\mathcal{L}$, which support a {\em retrieve} operation that returns the item at position $z$ in $\Delta \Q(\R,t)$ for any given $z$.  The index should also be able to return $N_B$, the size of the batch.  We further maintain a variable $\textsf{pos}$ to indicate the position of the current item retrieved.  Then the three primitives required by the batched reservoir sampling algorithm can be implemented as follows: 
\begin{itemize}[leftmargin=*]
    \item {\bf \remain()} returns $N_B - \textsf{pos}$, where $N_B$ is the size of batch;
    \item {\bf \sskip(i)} increases $\pos$ by $i + 1$ and returns the item at $\pos$ \xiao{(i.e., skips the next $i$ items and jumps directly to the $(i+1)$-th item)}; 
    \item {\bf \nnext()} simply returns {\bf \sskip(0)}; 
\end{itemize} 

Thus, to apply batched reservoir sampling on any join query $\Q$, it suffices to show how to maintain a linear-size index $\mathcal{L}$ that can efficiently support the retrieve operation for each $\Delta \Q(\R, t)$, as well as $|\Delta \Q(\R, t)|$.  This is still hard.  \xiao{To get around this difficulty, in Section~\ref{sec:acyclic} we devise an approximate solution.  Our index  $\mathcal{L}$ will implicitly define a $\Delta J$ that contains all the join results in  $\Delta \Q(\R, t)$, plus some dummy results.  However, we should not sample from these dummy join results, and this is exactly the reason why we must use a reservoir sampling algorithm that supports predicate. 
We set the predicate $\theta$ to \textsf{isReal}($\cdot$), which filters out the dummy results. We conceptually add some dummy tuples to base relations as well as some dummy partial join results. In this way, a join result is real if and only if all participated tuples are real, and dummy otherwise (i.e., at least one participated tuple or partial join result is dummy). The details of these dummy join results will be clear in Section~\ref{sec:line-3} and Section~\ref{sec:acyclic-join}.
} Finally, we will also guarantee that each $\Delta J$ is dense so as to apply Theorem \ref{the:batch-reservoir}.

\paragraph{Comparison with~\cite{zhao2020efficient}} The algorithm in \citet{zhao2020efficient} in fact follows the same framework, but they simply used the classical reservoir sampling algorithm without predicate.  As such, they must use an index $\mathcal{L}$ that supports the retrieve operation and the size information directly on $\Delta \Q(\R,t)$; please see Figure~\ref{fig:framework} for an illustration.  Such an index takes $O(N)$ time to update, although they used some heuristics to improve its practical performance. On the other hand, our predicate-enabled reservoir sampling algorithm allows us to use an $\mathcal{L}$ filled with dummy join results, which can be updated in $O(\log N)$ time as shown in the next section. 

\section{Sampling over Acyclic Joins}
\label{sec:acyclic}

We first recall the definition of an acyclic join:
\begin{definition}[Acyclic Join~\cite{beeri1983desirability}]
\label{def:acyclic}
    A (natural) join $\Q = (\V, \E)$ is $\alpha$-acyclic if there exists a tree $\T$ (called the join tree) such that (1) there is a one-to-one correspondence between the relations in $\E$ and nodes in $\T$; and (2) for every attribute $X \in \V$, the set of nodes in $\T$ containing $X$ are connected in $\T$. 
\end{definition} 

In this section, we present a dynamic index that can achieve the following guarantees:

\begin{theorem}
\label{the:acyclic}
Given any acyclic join $\Q$, an initially empty database $\R$, and a stream of $N$ tuples, we can maintain an index $\mathcal{L}$ on $\R$ \xiao{using $O(N)$ space} while supporting the following operations:
\begin{enumerate}[leftmargin=*]
    \item After a tuple $t$ is added to $\R$, $\mathcal{L}$ can be updated in $O(\log N)$ time amortized.
    \item The index implicitly defines an array $J \supseteq \Q(\R)$ where the tuples in $\Q(\R)$ are the real tuples and the others are dummy. 
    The index can return $|J|$ in $O(1)$ time.  For any given $j\in [|J|]$, it can return $J[j]$ in $O(\log N)$ time. Furthermore, $J$ is guaranteed to be $\phi$-dense for some constant $0<\phi\le 1$.
    \item The above is also supported for the delta query $\Delta \Q(\R, t)$ for any $t\not\in \R$.
\end{enumerate}
\end{theorem}

This index (using operation (1) and (2) above) immediately solves the dynamic sampling over join problem with an update time of $O(\log N)$ and sampling time $O(\log N)$.  Thanks to operation (3) and the density guarantee, it also solves the reservoir sampling over join problem by plugging into Theorem \ref{the:batch-reservoir} with $\alpha=\gamma=O(\log n)$ and $\beta = O(1)$.  The number of batches is $m=N$, while the $N$ in Theorem \ref{the:batch-reservoir}, which corresponds to the stream of join results now, becomes $N^{\rho^*}$.  Note that $\rho^*$ only depends on the query and not the input size, so it is taken as a constant. 

\begin{corollary}
Given any acyclic join $\Q$, an initially empty database $\R$, a sample size $k$, and a stream of $N$ tuples, \binyang{Algorithm \ref{alg:reservoir-sampling-join}} maintains $k$ uniform samples without replacement for each $\Q(\R^i)$, and runs in $O(N \log N + k \log N \log \frac{N}{k})$ expected time.
\end{corollary}

In fact, operation (2) can be reduced to operation (3): The $J$ for $\Q(\R)$ is just the concatenation of all the $\Delta J$'s of the delta queries.  The size of $J$ is the sum of all the $|\Delta J|$'s, which can be easily maintained in $O(1)$ time.  The concatenated $J$ is still dense as long as each $\Delta J$ is dense, due to Lemma \ref{lem:dense_concat}.  Henceforth we will only focus on operation (1) and (3). 

\subsection{Two-table Join}

We start by considering the simple two-table join $R_1(X,Y) \Join R_2(Y,Z)$.  For this query, the index simply consists of two arrays $R_1 \ltimes b$ and $R_2 \ltimes b$, as well as their sizes, for every $b \in \dom(Y)$. \xiao{The size of two arrays $R_1 \ltimes b$ and $R_2 \ltimes b$ is $|R_1 \ltimes b|$ and $|R_2\ltimes b|$ respectively. Summing over all $b\in \dom(Y)$, the whole index uses $O(N)$ space.} Then operation (1) can be easily supported in $O(1)$ time: We just add the tuple $t$ to $R_1 \ltimes t.Y$ if $t\in R_1$, or $R_2 \ltimes t.Y$ if $t\in R_2$. For operation (3), suppose $t\in R_1$.  We set $\Delta J = R_2 \ltimes t$.  Clearly, $J$ is $1$-dense as there are no dummy tuples, and any $J[j]$ can be retrieved in $O(1)$ time.

\subsection{Line-3 Join}
\label{sec:line-3}
When we move to the line-3 join $R_1(X,Y) \Join R_2(Y,Z) \Join R_3(Z,W)$, the situation becomes more complicated. However, even maintaining an index for just finding the delta query sizes is difficult: It is still an open problem if there is a better algorithm than computing each delta query size from scratch, which takes $O(N)$ time.  This is where we need to introduce dummy join results.

\paragraph{Index Structure} For each $b \in \pi_Y R_1$, we maintain the degree of $b$ in $R_1$, i.e., $\cnt(b) = |R_1 \ltimes b|$ and its approximation $\wcnt(b) = 2^{\lceil \log_2 \cnt(b)\rceil}$ by rounding $\cnt(b)$ up to the nearest power of $2$. Similarly, we maintain $\cnt(c) = |R_3 \ltimes c|$ and $\wcnt(c) = 2^{\lceil \log_2 \cnt(c)\rceil}$ for each $c \in \pi_{Z} R_3$. Note that $\wcnt(\cdot)$ changes at most $O(\log N)$ times.

\begin{figure*}
    \centering
    \includegraphics[scale=0.67]{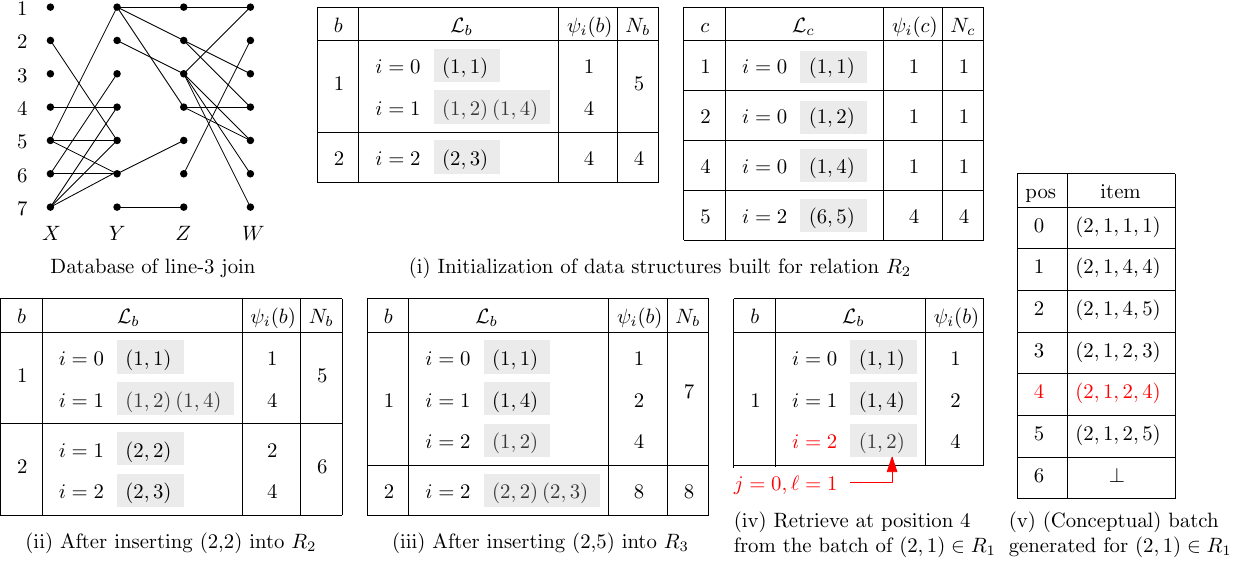}
    \vspace{-1em}
    \caption{An illustration of index structure built for line-3 join $R_1(X,Y) \Join R_2(Y,Z)\Join R_3(Z,W)$.}
    \label{fig:line3}
\end{figure*}

For each value $b \in \pi_Y R_2$, we organize the tuples $R_2 \ltimes b$ into at most $\log N$ buckets according to the approximate degree of $c$, where the $i$-th bucket is
\[\Phi_i(b) = \left\{(b,c) \in R_2: \wcnt(c) = 2^i \right\}.\]
Let $\mathcal{L}_b$ be the list of non-empty buckets. 
Define $\varphi_i(b) = 2^{i} \cdot |\Phi_{i}(b)|$.  We also maintain $\nsum_b = \sum_{i \in [\log N]} \varphi_i(b)$ for each value $b \in \pi_Y R_2$, which is an upper bound on the number of new join results if some tuple $(a,b)$ is added to $R_1$. Symmetrically, for each $c \in \pi_Z R_2$, we maintain such a list $\mathcal{L}_c$, and $\nsum_c = \sum_{i \in [\log N]} \varphi_i(c)$.   Please see Figure~\ref{fig:line3} for an example.

\xiao{\paragraph{Space Usage} As there are $O(N)$ values in $\pi_Y R_1$, we need to maintain $O(N)$ degrees and their approximations in total. For each $b \in \pi_Y R_2$, it needs to organize the tuples $R_2\ltimes b$ into buckets and maintain a value $\nsum_b$. The size of non-empty buckets maintained for $b$ is essentially $|R_2 \ltimes b|$. Summing over all values $b \in \pi_Y R_2$, the total size is $O(N)$. Similar argument applies to $\pi_{Z} R_2$.}

\paragraph{Index Update} After a tuple $t$ has arrived, we update our data structure as follows.  If $t \in R_2$, say $t= (b,c)$, we add $(b,c)$ to $\Phi_i(b)$ for $i = \log_2 \wcnt(c)$. This just takes $O(1)$ time. 

If $t = (a,b) \in R_1$ (the $t\in R_3$ case is similar), we increase $\cnt(b)$ by 1, and update $\wcnt(b)$ if needed. If $\wcnt(b)$ has changed, for each $c \in \pi_{Z} \left(R_2\ltimes b\right)$, we remove $(b,c)$ from $\Phi_{i-1}(c)$ and add $(b,c)$ to $\Phi_i(c)$, where \binyang{$i = \log_2 \wcnt(b)$}.  This may take $O(N)$ time, but this update is only triggered when $\wcnt(b)$ doubles, which happens at most $O(\log N)$ times. Thus, the total update cost is 
\begin{align*}
    \sum_{b} \lceil \log \cnt(b) \rceil \cdot |\pi_{Z} \left(R_2 \ltimes b\right)| &\le \log N \cdot \sum_{b}|\pi_{Z} \left(R_2 \ltimes b\right)|\le N \log N,
\end{align*}
namely, the amortized update cost is $O(\log N)$.  Finally, whenever some $\Phi_i(b)$ or $\Phi_i(c)$ changes, we update $\nsum_b$ and $\nsum_c$ accordingly. The time for this update is the same as that for $\Phi_i(b)$ and $\Phi_i(c)$.

\paragraph{Batch Generate}
The delta query $\Delta \Q(\R,t)$ on the line-3 join falls into the following 3 cases:
\begin{displaymath}
\Delta \Q(\R,t) = \left\{ \begin{array}{ll}
\{t\} \times (R_3 \ltimes (R_2 \ltimes b)) & \textrm{if $t =(a,b)\in R_1$}\\
(R_1 \ltimes b) \times \{t\} \ltimes (R_3 \ltimes c) & \textrm{if $t=(b,c) \in R_2$}\\
(R_1 \ltimes (R_2 \ltimes c)) \times \{t\} & \textrm{if $t =(c,d)\in R_3$}
\end{array} \right.
\end{displaymath}

The batch $\Delta J \supseteq \Delta\Q(\R,t)$ for any $t$ is defined as follows.
If $t \in R_2$, say $t = (b,c)$, then $\Delta J := (R_1 \ltimes b) \times (R_3 \ltimes c)$. This batch is $1$-dense and $|\Delta J| = \cnt(b) \cdot \cnt(c)$.

Next, consider the case $t=(a,b)\in R_1$. 
Consider a bucket $\langle i, \Phi_i(b)\rangle \in \mathcal{L}_b$. For each $(b,c) \in \Phi_i(b)$, define a mini-batch that consists of all tuples in $R_3 \ltimes c$, followed by $\wcnt(c) - \cnt(c)$ dummy tuples. We concatenate these mini-batches to form the batch for the bucket, and concatenate all the buckets to form $\Delta J$.  This $\Delta J$ is ${1 \over 2}$-dense, since each mini-batch is ${1\over 2}$-dense and then we invoke Lemma~\ref{lem:dense_concat}.  
Moreover, $|\Delta J| = \nsum_b$ and can be returned in $O(1)$ time. The case $t\in R_3$ is similar.

\paragraph{Retrieve} 
We next show how to retrieve a specific element from the $\Delta J$ defined above. We  consider the two cases $t\in R_2$ and $t\in R_1$ ($t \in R_3$ is similar), respectively.

If $t =(b,c)\in R_2$, $\Delta J$ is the Cartesian product of $R_2 \ltimes b$ and $R_3 \ltimes c$. Given a position $z \in [|\Delta J|]$, we first find the unique pair $(z_1,z_2) \in [|R_2 \ltimes b|] \times [|R_3 \ltimes c|]$ such that $z = z_1 \cdot |R_3 \ltimes c| + z_2$. Then, we just return the combination of the tuple at position $z_1$ in $R_2 \ltimes b$ and the tuple at position $z_2$ in $R_3 \ltimes c$. The retrieve operation in this case takes  $O(1)$ time.

If $t =(a,b) \in R_1$, we retrieve the tuple at position $z$ as follows: 
\begin{itemize}[leftmargin=*]
    \item Let $i \in [0,\log N]$ be the unique integer such that \[\sum_{i' \le i-1: \Phi_{i'}(b) \neq \emptyset} \varphi_{i'}(b) < z + 1 \le \sum_{i' \le i: \Phi_{i'}(b) \neq \emptyset} \varphi_{i'}(b).\]
    \item Set $\displaystyle{j  = \left \lfloor (z - \sum_{i' \le i-1: \Phi_{i'}(b) \neq \emptyset} \varphi_{i'}(b))/2^i \right \rfloor}$.
    \item Set $\displaystyle{\ell  = z - \sum_{i' \le i-1: \Phi_{i'}(b) \neq \emptyset} \varphi_{i'}(b) - 2^i \cdot j}$.
\end{itemize}
Let $t'$ be the tuple at position $j$ in $\Phi_i(b)$. Then we return the tuple at position $\ell$ in $R_3 \ltimes t'$ if $\ell < |R_3 \ltimes t'|$, and a dummy tuple otherwise.  As there are at most $O(\log N)$ distinct $i$'s with $\Phi_i(b) \neq \emptyset$, the value of $i,j,\ell$ can be computed in $O(\log N)$ time.  So the retrieve operation takes $O(\log N)$ time in this case.

\subsection{Acyclic Joins}
\label{sec:acyclic-join}
Finally, we generalize the line-3 algorithm to an arbitrary acyclic join  $\Q = (\V,\E)$.  Let $\T$ be any join tree of $\Q$. Recall that there is a one-to-one correspondence between nodes in $\T$ and relations in $\E$. Moreover, for every attribute $X \in \V$, all nodes containing $X$ are connected in $\T$. By definition, $\T$ is an unrooted tree, but we can root it by specifying any node as the root $r$.  We will  consider all the rooted trees where $r$ ranges over all nodes, and the one with root $r$ will be responsible for generating the batch $\Delta J \supseteq \Delta \Q(\R, t)$ for any $t\in R_r$.  For example, the line-3 join has one unrooted join tree $R_1-R_2-R_3$ but 3 rooted trees.  The line-3 algorithm can be conceptually considered as 3 algorithms, each using one rooted tree.  Some data structures among them can be shared, but below we will just focus on one rooted tree for conceptual simplicity. 

Consider a $\T$ rooted at $r$.  We use $p_e$ to denote the parent of $e$. For the root $r$, set $p_r = \emptyset$. Let $\key(e) = e \cap p_e$ be the common attributes shared between $e$ and its parent $p_e$. Let $\mathcal{C}_e$ be the child nodes of node $e$. For a leaf node $e$, $\mathcal{C}_e = \emptyset$. Let $\T_e$ be the sub-tree below $e$. With a slight abuse of notation, we also use $\T_e$ to denote the set of relations whose corresponding nodes are in $\T_e$. 

\paragraph{Index structure} We store input tuples in a hash table, so that for any $e \in \E$, a subset of attributes $x \subseteq e$ and a tuple $t \in \dom(x)$, we can get the list of tuples $R_e \ltimes t$ in $O(1)$ time.  For each node $e \in \T$ and tuple $t \in \pi_{\key(e)} R_e$, we maintain a upper bound $\cnt[\T, e, t]$ on the {\em degree} of $t$ in $\T_e$, i.e., the number of join results over relations in $\T_e$ whose projection onto attributes $\key(e)$ matches $t$:
\begin{displaymath}
\cnt[\T, e,t]  = \left\{ \begin{array}{ll}
|R_e \ltimes t| & \textrm{if $e$ is a leaf}\\
\displaystyle{\sum_{t' \in R_e \ltimes t} \prod_{e' \in \mathcal{C}_e} \wcnt\left[\T, e', \pi_{\key(e')} t' \right]} & \textrm{otherwise}
\end{array} \right.
\end{displaymath}

Note that this definition depends on $\wcnt(\cdot)$, which is recursively defined as $\wcnt[\T, e,t] = 2^{\lceil \log_2 \cnt[\T,e,t] \rceil}$ by rounding $\cnt[\T,e,t]$ up to the nearest power of $2$. We point out an important property of $\wcnt(\cdot)$ in Lemma~\ref{lem:wcnt}, which indicates that $\wcnt[\T,e,t]$ is a constant-approximation of the 
degree of $t$ in $\T_e$. 
\begin{lemma}
\label{lem:wcnt}
    For a join tree $\T$, node $e$ and tuple $t \in \pi_{\key(e)} R_e$,
    $\displaystyle{\wcnt[\T, e, t] \le 2^{|\T_e|} \cdot \left|\left(\Join_{e' \in \T_e} R_{e'}\right) \ltimes t\right|}$.
\end{lemma}

\begin{proof}
    We prove it by induction. If $e$ is a leaf node, $\cnt[\T, e, t] = \left|R_e \ltimes t\right|$. As $\wcnt[\T, e,t] \le 2 \cnt[\T, e,t]$, we have $\wcnt[\T, e, t] \le 2 \cdot \left|R_e \ltimes t \right|$. If $e$ is an internal node, we assume the lemma holds for every child node $e' \in \mathcal{C}_e$ and tuple $t' \in \pi_{\key(e')} R_{e'}$. For an arbitrary tuple $t \in \pi_{\key(e)} R_e$, we can bound $\wcnt[\T,e,t]$ as
    \begin{align*}
        \wcnt[\T,e, t] \le 2 \cdot \cnt[\T,e,t] 
        = & 2 \cdot \sum_{t' \in R_e \ltimes t} \prod_{e' \in \mathcal{C}_e} \wcnt\left[\T, e', \pi_{\key(e')} t' \right] \\
        \le & 2 \cdot \sum_{t' \in R_e \ltimes t} \prod_{e' \in \mathcal{C}_e} 2^{|\T_{e'}|} \cdot \left|\left(\Join_{e'' \in \T_{e'}} R_{e''}\right)\ltimes \left(\pi_{\key(e')} t' \right)\right|\\
        = & 2 \cdot 2^{|\T_e| -1} \cdot \left|\left(\Join_{e' \in \T_e}R_{e'}\right) \ltimes t \right|
    \end{align*}
    where the last equality follows the intersections property of $\T$. 
\end{proof}

Together with the fact that $\left|\left(\Join_{e' \in \T_e} R_{e'}\right) \ltimes t\right| \le \left|\Join_{e' \in \T_e} R_{e'}\right| \le N^{|\T_e|}$, we obtain $\wcnt[\T, e, t] \le (2N)^{|\T_e|}$, which implies that $\wcnt[\T, e, t]$ can only be doubled by at most $O(\log N)$ times. 

Consider any non-root node $e$. For each tuple $t \in \pi_{\key(e)} R_e$, we organize the tuples $t' \in R_e \ltimes t$
into at most $|\T_e| \cdot \log 2N$ buckets according to the approximate degree of $t'$ in $\T_e$, where the $i$-th bucket is
\[\Phi_{i,e}(t) = \left\{t' \in R_e \ltimes t: \prod_{e' \in \mathcal{C}_e}\wcnt\left[\T, e',\pi_{\key(e')} t'\right] = 2^i\right\}.\]
Let $\mathcal{L}_{e,t}$ be the list of non-empty buckets. For simplicity, we denote $\varphi_{i,e}(t) = 2^{i} \cdot |\Phi_{i,e}(t)|$ for each $i \in [|\T_e| \cdot \log 2N]$. We also maintain $\nsum_t = \sum_{i \in [|\T_e| \cdot \log 2N]} \varphi_{i,e}(t)$ for each $t \in \pi_{\key(e)} R_e$.

\paragraph{\binyang{Space Usage}} \binyang{We consider an arbitrary $e \in \E$ in an arbitrary join tree maintained. Let $\mathcal{C}_e$ be the children of $R_e$. We build an index on $R_e$ with $\key(e_i)$ as the key for each $e_i \in \mathcal{C}_e$ in order to perform the look up in line 9 of Algorithm~\ref{alg:index-update}. There are $|\mathcal{C}_e|$ such indices of size $O(N)$ in total. 
Moreover, for any non-root node $e$ and each tuple $t \in \pi_{\key(e)} R_e$, we organize tuples $R_e \ltimes t$ into buckets and maintain $\nsum_t$. All these buckets are disjoint and the total size is $O(N)$. As there are $O(1)$ join trees and each join tree contains $O(1)$ nodes, the whole index uses $O(N)$ space.}

\paragraph{Index Update} We define a generalized procedure for updating our index. As described in Algorithm~\ref{alg:index-update}, the procedure \textsc{IndexUpdate} takes as input the join tree $\T$, a node $e$, tuple $t \in R_e$ and an integer $\textsf{old} \ge 0$ (indicating the approximated degree of $t$ in $\T_e$ before update). The update proceeds in a recursive way. We first compute the approximated degree of $t$ in $\T_e$ after update, denoted as $\textsf{new}$. For simplicity, denote $t_e = \pi_{\key(e)}$. We then remove $t$ from the old bucket if it exists (line 4) and insert $t$ into the new bucket (line 5). 
We increase $\cnt[\T,e, t_e]$ by $\textsf{new} - \textsf{old}$ (line 7), and update $\wcnt[\T,e, t_e]$ if needed. If $\wcnt[\T,e,t_e]$ has changed, we might need to propagate the updates upward (line 8-11). Furthermore, if $p_e$ is not the root, for each tuple $t' \in R_{p_e} \ltimes t_e$, we compute the approximated degree of $t'$ in $\T_{p_e}$ before update (line 10), denoted as $\textsf{old}'$ and invoke this whole procedure recursively (line 11).

\begin{algorithm}[t]
\caption{$\textsc{IndexUpdate}(\T,e,t, \textsf{old})$}
\label{alg:index-update}

\SetKwInOut{Input}{Input}
\SetKwInOut{Output}{Output}

\Input{A join tree $\T$ for $\Q$, a node $e$ and tuple $t \in R_e$, an approximate degree $\textsf{old}$ of $t$ in $\T_e$ before update;}
\Output{Updated $\cnt(\cdot )$ and $\wcnt(\cdot)$;}

$t_e \gets \pi_{\key(e)} t$\;
$\textsf{new} \gets \displaystyle{\prod_{e' \in \mathcal{C}_e} \wcnt[\T, e', \pi_{\key(e')} t]}$\;
$i' \gets \log_2 \textsf{old}$ and $i \gets \log_2 \textsf{new}$\;
\lIf {$i' > 0$}{$\Phi_{i', e}(t_e) \gets \Phi_{i',e}(t_e) - \{t\}$}

$\Phi_{i, e}(t_e) \gets \Phi_{i,e}(t_e) \cup \{t\}$\;
$j \gets 2^{\lceil \log \cnt\left[\T, e,t_e \right]\rceil}$\;
$\displaystyle{\cnt\left[\T, e, t_e \right]\gets  \cnt\left[\T, e, t_e\right] + 
 \textsf{new}} - \textsf{old}$\;
\If{$\wcnt\left[\T, e,t_e \right]$ changes and $p_e$ is not the root}{
    \ForEach{$t' \in R_{p_e} \ltimes t_e$}{
        $\textsf{old}' \gets \displaystyle{j \cdot \prod_{e' \in \mathcal{C}_{p_e} - \{e\}} \wcnt[\T, e', \pi_{\key(e')} t']}$\;
        $\textsc{IndexUpdate}(\T,p_e,t',\textsf{old}')$\;
    }
}
\end{algorithm}

When a tuple $t$ is inserted into $R_e$, we just invoke $\textsc{IndexUpdate}$ $\left(\T, e, t, 0 \right)$ for every join tree $\T$ used in our index. This may take $O(N)$ time, but the observation is that this update is only triggered when $\wcnt[\T, e, t_e]$ changes, which happens at most $O(\log N)$ times. Thus, the total update cost is:
\begin{align*}
    \sum_{e \in \T} \sum_{t \in \pi_{\key(e)} R_e} \log N \cdot \left| R_{p_e} \ltimes t \right| 
    \le &  \log N \cdot \sum_{e \in \T} \sum_{t \in \pi_{\key(e)} R_e} \left|R_{p_e} \ltimes t\right| \\
    \le  & \log N \cdot \sum_{e' \in \T: e' \textrm{ is an internal node }} |R_{e'}| \cdot |\mathcal{C}_{e'}|= O(N\log N).
\end{align*}
Whenever some $\Phi_{i,e}(t)$ changes, we update $\nsum_t$ accordingly.  The time for this update is the same as that for updating $\Phi_{i,e}(t)$. Finally, summing over all join trees used, each having a distinct relation as its root, the overall update cost is $O(N \log N)$, namely, the amortized update cost is $O(\log N)$.
\begin{algorithm}[t]
\caption{$\textsc{BatchGenerate}(\T,e,t)$}
\label{alg:batch-generate}
    \SetKwInOut{Input}{Input}
    \SetKwInOut{Output}{Output}

    \Input{A join tree $\T$ for $\Q$, a node $e$ and a tuple $t \in R_e$ or $t \in \pi_{\key(e)} R_e$;}
    \Output{A $O(1)$-dense batch $\Delta J \supseteq \Delta\Q(\R,t)$;}
    
    $x \gets \textsf{supp}(t)$\;
    \lIf{$e$ is a leaf node and $x =e$}{\Return $t$}
    \If{$e$ is an internal node and $x =e$}{
        \ForEach{$e_i \in \mathcal{C}_e$}{$B_i \gets \textsf{BatchGenerate}(\T, e_i, \pi_{\key(e_i)}t)$\;}
       \Return $\{t\} \times \left(\times_{e_i \in \mathcal{C}_e} B_i\right)$\; 
    }
        \lFor{$t' \in R_e \ltimes t$}{$B_{t'} \gets \textsf{BatchGenerate}(\T, e, t')$}
        $L \gets \wcnt[\T,e,t] - \cnt[\T,e,t]$ dummy elements\;
        \Return concatenation of $B_{t'}$ for $t' \in R_e \ltimes t$, followed by $L$\; 
\end{algorithm}
\begin{algorithm}[t]
\caption{$\textsc{Retrieve}(\T,e,t,z)$}
\label{alg:retrieve}

    \SetKwInOut{Input}{Input}
    \SetKwInOut{Output}{Output}

    \Input{A join tree $\T$ for $\Q$, a node $e$ and \xiao{a tuple} $t\in R_e$ or $t \in \pi_{\key(e)} R_e$, an integer $z\ge 0$;}
    \Output{The element at position $z$ in the batch generated for $t$ by \textsc{BatchGenerate}$(\T,e,t)$;}
    
    $x \gets \textsf{supp}(t)$\;
    \If{$e$ is a leaf node}{
        \lIf{$z \ge \cnt[\T,e,t]$}{\Return $\perp$}
        \lElse{\Return the element at position $z$ in $R_e \ltimes t$}
        }
    \uIf{$e =x$}{
        $\mathcal{C}_e \gets \{e_1, e_2, \cdots, e_m\}$\;
        \lForEach{$i \in [1\dots m]$}{$t_i \gets \pi_{\key(e_i)}t$}
        Find $(z_1,z_2,\cdots,z_m) \in \times_{i=1}^m \wcnt[\T, e_i,t_i]$ such that $\displaystyle{z = \sum_{i \in [1\dots m]} \left(z_i \cdot \prod_{j > i} \wcnt[\T, e_j, t_j]\right)}$\;
        \ForEach{$i \in [1\dots m]$}{
            $t'_i \gets \textsc{Retrieve}(\T, e_i, t_i, z_i)$\;
            \lIf{$t'_i = \perp$}{\Return $\perp$}
        }
        \Return $(t'_1, t'_2, \cdots, t'_m)$\;
    }
    \Else{
        \lIf{$z \ge \cnt[\T,e,t]$}{\Return $\perp$}
        Find $i$ such that $\displaystyle{\sum_{i' \le i-1} \varphi_{i',e}(t) < z + 1 \le \sum_{i' \le i} \varphi_{i',e}(t)}$\;
        $\displaystyle{j  \gets \left \lfloor \left(z - \sum_{i' \le i-1} \varphi_{i',e}(t)\right)/2^i \right \rfloor}$\;
        $\displaystyle{\ell \gets  z - \sum_{i' \le i-1} \varphi_{i',e}(t) - 2^i \cdot j}$\; 
        $t' \gets$ the element at position $j$ in $\Phi_{i,e}(t)$\;        
        \Return $\textsc{Retrieve}(\T, e, t',\ell)$\;  
    }
\end{algorithm}

\paragraph{Batch Generate} We define a generalized procedure for generating an $\Omega(1)$-dense batch $\Delta J \supseteq \Delta \Q(\R,t)$ for any tuple $t\in R_e$ or $t \in \pi_{\key(e)} R_e$, as described in Algorithm~\ref{alg:batch-generate}. The density will depend on the query size, which is taken as a constant, but not on the data size.
If tuple $t$ is inserted into $R_e$, the fist call is $\textsc{BatchGenerate}(\T,e,t)$, where $\T$ is the join tree rooted at node $e$. In each recursive call, Algorithm~\ref{alg:batch-generate} distinguishes three cases:
\begin{itemize}[leftmargin=*]
    \item {\bf Case 1:} $e$ is a leaf node and $t\in R_e$. We simply return $t$ as $\Delta J$. This batch is $1$-dense and $|\Delta J|=1$.
    \item {\bf Case 2:} $e$ is an internal node and $t \in R_e$. In this case, $\Delta \Q(\R,t)$ can be decomposed into the Cartesian product of $\Delta \Q(\R, \pi_{\key(e_i)}t)$ for each child $e_i \in \mathcal{C}_e$. The batch $\Delta J$ also follows the same way. We recursively generate a batch for $\pi_{\key(e_i)} t$ in $\T_{e_i}$ for each $e_i \in \C_e$, and return their cross product as $\Delta J$. 
    \item {\bf Case 3:} $t \in \pi_{\key(e)} R_e$. We recursively generate a batch for every tuple $t' \in R_e \ltimes t$ in $\T_e$ and concatenate these batches with $\wcnt[\T,e,t] - \cnt[\T,e,t]$ dummy elements at the end as $\Delta J$. 
\end{itemize} 

It can be easily shown by induction that 
\begin{displaymath}
|\Delta J| = \left\{ \begin{array}{ll}
\displaystyle{\prod_{e_i\in \mathcal{C}_e} \wcnt[\T,e_i,\pi_{\key(e_i)}t]} & \textrm{if $t \in R_e$ } \\
\displaystyle{\wcnt[\T,e,t]} & \textrm{if $t \in \pi_{\key(e)} R_e$} 
\end{array} \right.
\end{displaymath}
where the second case follows the definition of $\cnt(\cdot)$. Hence, $|\Delta J|$ can be returned in $O(1)$ time. We next prove by induction that 
\begin{displaymath}
\textrm{$\Delta J$ is $\phi$-dense, where } \phi = \left\{ \begin{array}{ll}
(\frac{1}{2})^{2 \cdot |\T_e|-1} & \textrm{if $t \in \pi_{\key(e)} R_e$} \\ (\frac{1}{2})^{2 \cdot |\T_e| - 2} & \textrm{if $t \in R_e$ } 
\end{array} \right.
\end{displaymath}
This holds trivially for {\bf Case 1}. For {\bf Case 2}, we assume that the batch generated for tuple $\pi_{\key(e_i)} t$ is $(\frac{1}{2})^{2\cdot |\T_{e_i}|-1}$-dense. Implied by Lemma~\ref{lem:dense_cross_product}, this $\Delta J$ is $(\frac{1}{2})^{2 \cdot |\T_{e}|-2}$-dense, since
\[ (\frac{1}{2})^{|\C_e|-1} \cdot \prod_{e_i \in \C_e}(\frac{1}{2})^{2 \cdot |\T_{e_i}|-1} = (\frac{1}{2})^{2 \cdot |\T_{e}| - 3} \ge (\frac{1}{2})^{2 \cdot |\T_{e}| - 2}\]
For {\bf Case 3}, we assume the batch generated for each tuple $t' \in R_e \ltimes t$ is $(\frac{1}{2})^{2\cdot|\T_e| - 2}$-dense. Their concatenation is also $(\frac{1}{2})^{2\cdot|\T_e| - 2}$-dense, implied by Lemma~\ref{lem:dense_concat}. The $\Delta J$ is $(\frac{1}{2})^{2\cdot |\T_e| - 1}$-dense, since $\cnt[\T,e,t] \ge \frac{1}{2} \cdot \wcnt[\T,e,t]$ and then we invoke Lemma~\ref{lem:dense_padding}.

\smallskip \noindent {\bf Retrieve.} Finally, we describe how to retrieve the join result at position $z$ in the batch generated by \textsc{BatchGenerate}. As described in Algorithm~\ref{alg:retrieve}, \textsc{Retrieve} follows the same recursive structure as that of \textsc{BatchGenerate}. If tuple $t$ is inserted into $R_e$, the first call is $\textsc{Retrieve}(\T,e,t,z)$, where $\T$ is join tree rooted at node $e$. In each recursive call, we also distinguish three cases:
\begin{itemize}[leftmargin=*]
    \item {\bf Case 1:} $e$ is a leaf node. We simply return the element at position $z$ in the batch accordingly. This takes $O(1)$ time. 
    \item {\bf Case 2:} $e$ is an internal node and $t \in R_e$. In this case, we decompose the index $z$ into a $m$-coordinate $(z_1,z_2,\cdots, z_m)$ as defined in line 8, then retrieve the element at position $z_i$ in the batch generated for tuple $\pi_{\key(e_i)} t$ for each $e_i \in \mathcal{C}_e$ recursively (line 9), and return their combinations as the final result (line 12). The value of $z_1, z_2,\cdots, z_m$ can be computed in $O(1)$ time. 
    \item {\bf Case 3:} $e$ is an internal node and $t \in \pi_{\key(e)} R_e$. In this case, we first locate the bucket into which the element at position $z$ falls, say $i$. We then locate the index of the tuple whose batch contains the element at position $z$, say $j$, and find the specific tuple $t'$. We also need to compute the index of the target element in the batch generated for $t'$, say $\ell$. Finally, the element at position $z$ in the batch generated for $t$ can be found by $\textsc{Retrieve}(\T, e, t', \ell)$. The value of $i,j,\ell$ can be computed in $O(\log N)$ time.
\end{itemize} 
It is not hard to see that the retrieve operation takes $O(\log N)$ time, by summing the time cost for each recursive invocation.

\subsection{Optimizations}
\label{sec:optimization}
We next discuss some optimization techniques for our algorithm.  Although they do not improve the  complexity results, they significantly reduce the constant factor, as verified in Section~\ref{sec:experiment}. 

\paragraph{Grouping}
In a join tree $\T$, for an non-root internal node $R_e$ with its children nodes $\{e_1,e_2,\cdots,e_m\}$, let $\bar{e} = \key(e) \cup \key(e_1) \cup \cdots \cup \key(e_m)$ denote the join attributes. We can {\em group} tuples in $R_e$ by attributes $\bar{e}$, if $e - \bar{e}\neq \emptyset$. Let $R_{\bar{e}} = \pi_{\bar{e}} R_e$. We replace $e$ with $\bar{e}$ in $\T$. For each tuple $t \in R_{\bar{e}}$, we maintain $\mul[\T,\bar{e},t] = |R_e \ltimes t|$ and $\wmul[\T,\bar{e},t] = 2^{\lceil \log_2 \mul[\T,\bar{e},t] \rceil}$. 
Then, $\cnt[\T, \bar{e}, t]$ is defined as:
\[\displaystyle{\cnt[\T,\bar{e},t] = \sum_{t' \in R_{\bar{e}} \ltimes t} \wmul[\T,\bar{e},t'] \cdot \prod_{e_i \in \C_{\bar{e}}}\wcnt\left[\T, e_i, \pi_{\key(e_i)} t' \right]}\] 

The grouping version of update and retrieve procedures are described in Algorithm~\ref{alg:index-update-with-grouping} and Algorithm~\ref{alg:retrieve_with_grouping}.
For the tuple $t_e$ inserted into $R_e$, if $e - \bar{e} \neq \emptyset$, we first compute $t_{\bar{e}} = \pi_{\bar{e}}t$, and then \[\textsf{old} = \wmul[\T, \bar{e}, t_{\bar{e}}] \cdot \prod_{w \in \mathcal{C}_e} \wcnt[\T,w,\pi_{\key(w)}t_e]\]
\noindent We increase $\mul[\T, \bar{e}, t_{\bar{e}}]$ by 1 and update $\wmul[\T, \bar{e}, t_{\bar{e}}]$ accordingly. Finally, if $\wmul[\T, \bar{e}, t_{\bar{e}}]$ changes, we invoke $\textsc{IndexUpdateGrouping}$ $\left(\T, \bar{e},t_{\bar{e}}, \textsf{old} \right)$ to handle propagated updates recursively.
\begin{algorithm}[t]
\caption{$\textsc{IndexUpdateGrouping}(\T,e,t, \textsf{old})$}
\label{alg:index-update-with-grouping}
\SetKwInOut{Input}{Input}
\SetKwInOut{Output}{Output}

\Input{A join tree $\T$ for $\Q$, a node $e$ and tuple $t \in R_e$, an approximate degree $\textsf{old}$ of $t$ in $\T_e$ before update;}
\Output{Updated $\cnt(\cdot )$ and $\wcnt(\cdot)$;}

$t_e \gets \pi_{\key(e)} t$\;
$f \gets 1$\;
\lIf{$e$ is created by grouping}{$f \gets \wmul[\T,e,t]$}
$\textsf{new} \gets \displaystyle{f \cdot \prod_{e' \in \mathcal{C}_e} \wcnt[\T, e', \pi_{\key(e')} t]}$\;
$i' \gets \log_2 \textsf{old}$ and $i \gets \log_2 \textsf{new}$\;
\lIf {$i' > 0$}{$\Phi_{i', e}(t_e) \gets \Phi_{i',e}(t_e) - \{t\}$}

$\Phi_{i, e}(t_e) \gets \Phi_{i,e}(t_e) \cup \{t\}$\;
$j \gets \wcnt\left[\T, e,t_e \right]$\;
$\displaystyle{\cnt\left[\T, e, t_e \right]\gets  \cnt\left[\T, e, t_e\right] + 
 \textsf{new}} - \textsf{old}$\;
\If{$\wcnt\left[\T, e,t_e \right]$ changes and $p_e$ is not the root}{
    \ForEach{$t' \in R_{p_e} \ltimes t_e$}{   
        $f' \gets 1$\;
        \lIf{$p_e$ is created by grouping}{$f' \gets \wmul[\T,p_e,t']$}
        $\textsf{old}' \gets \displaystyle{j \cdot f' \cdot \prod_{e' \in \mathcal{C}_{p_e} - \{e\}} \wcnt[\T, e', \pi_{\key(e')} t']}$\;
        $\textsc{IndexUpdate}(\T,p_e,t',\textsf{old}')$\;}
}
\end{algorithm}
\begin{algorithm}
\caption{$\textsc{RetrieveWithGrouping}(\T,e,t,z)$}
\label{alg:retrieve_with_grouping}
     \SetKwInOut{Input}{Input}
    \SetKwInOut{Output}{Output}

    \Input{A join tree $\T$, a node $e$ and some tuple $t\in R_e$ or $t \in \pi_{\key(e)} R_e$, an integer $z\ge 0$;}
    \Output{The element at position $z$ in the batch generated for $t$ by \textsc{BatchGenerate}$(\T,e,t)$;}
    
    $x \gets \textsf{supp}(t)$\;
    \If{$e$ is a leaf node}{
        \lIf{$z \ge \cnt[\T,e,t]$}{\Return $\perp$}
        \lElse{\Return the element at position $z$ in $R_e \ltimes t$}
        }
    \If{$e =x$}{
        $\mathcal{C}_e \gets \{e_1, e_2, \cdots, e_m\}$\;
        \lForEach{$i \in [1\dots m]$}{$t_i \gets \pi_{\key(e_i)}t$}
        Find $(z_1,z_2,\cdots,z_m) \in \times_{i=1}^m \wcnt[\T, e_i,t_i]$ such that $\displaystyle{z = \sum_{i \in [1\dots m]} \left(z_i \cdot \prod_{j > i} \wcnt[\T, e_j, t_j]\right)}$\;
        \ForEach{$i \in [1\dots m]$}{
            $t'_i \gets \textsc{RetrieveWithGrouping}(\T, e_i, t_i, z_i)$\;
            \lIf{$t'_i = \perp$}{\Return $\perp$}
        }
        \Return $(t'_1, t'_2, \cdots, t'_m)$\;
    }
    \ElseIf{$e$ is created from $e'$ by grouping}{
            \lIf{$z \ge \cnt[\T,e,t]$}{\Return $\perp$}
            Find $i$ such that $\displaystyle{\sum_{i' \le i-1} \varphi_{i',e}(t) < z + 1 \le \sum_{i' \le i} \varphi_{i',e}(t)}$\;
            $\displaystyle{j  \gets \left \lfloor \left(z - \sum_{i' \le i-1} \varphi_{i',e}(t)\right)/2^i \right \rfloor}$\;
            $\displaystyle{\ell \gets  z - \sum_{i' \le i-1} \varphi_{i',e}(t) - 2^i \cdot j}$\; 
            $t' \gets$ the element at position $j$ in $\Phi_{i,e}(t)$\;  
            $\displaystyle{h \gets \prod_{e_i \in \mathcal{C}_e} \wcnt[\T,e_i,\pi_{\key(e_i)}t']}$\;
            \lIf{$\lfloor \ell/h \rfloor \ge \mul[\T,e,t']$}{\Return $\perp$}
            $f \gets \ell - \lfloor \ell / h \rfloor \cdot h$\;
            $t'' \gets$ the element at position $\lfloor \ell/h \rfloor$ in $R_{e'} \ltimes t'$\;
            \Return $\textsc{RetrieveWithGrouping}(\T, e, t'',f)$\;  
    }
    \Else{
        \lIf{$z \ge \cnt[\T,e,t]$}{\Return $\perp$}
        Find $i$ such that $\displaystyle{\sum_{i' \le i-1} \varphi_{i',e}(t) < z + 1 \le \sum_{i' \le i} \varphi_{i',e}(t)}$\;
        $\displaystyle{j  \gets \left \lfloor \left(z - \sum_{i' \le i-1} \varphi_{i',e}(t)\right)/2^i \right \rfloor}$\;
        $\displaystyle{\ell \gets  z - \sum_{i' \le i-1} \varphi_{i',e}(t) - 2^i \cdot j}$\; 
        $t' \gets$ the element at position $j$ in $\Phi_{i,e}(t)$\;        
        \Return $\textsc{RetrieveWithGrouping}(\T, e, t',\ell)$\; 
    }
\end{algorithm}
Grouping can bring much benefit in index update. More specifically, in line 9-11 of Algorithm~\ref{alg:index-update}, instead of propagating update for every tuple $t' \in R_{p_e} \ltimes t_e$, we now propagate update only for every tuple $t \in R_{\bar{p_e}} \ltimes t_e$, where $R_{\bar{p_e}}$ is the projection of relation $R_{p_e}$ onto join attributes in $p_e$. \xiao{Hence, we can see a significant reduction in the number of propagated updates. }

\begin{figure}
    \centering
    \begin{tabular}{llcccc}
$Ra$                         &                       & $R_b$                         & $R_{\bar{b}}$                                         & $\mul$                                     & $\wmul$                                    \\ \cline{1-1} \cline{3-6} 
\multicolumn{1}{|l|}{$(1,1)$} & \multicolumn{1}{l|}{} & \multicolumn{1}{c|}{$(1,1,2)$} & \multicolumn{1}{c|}{\multirow{3}{*}{$(1,2)$}} & \multicolumn{1}{c|}{\multirow{3}{*}{$3$}} & \multicolumn{1}{c|}{\multirow{3}{*}{$4$}} \\ \cline{1-1} \cline{3-3}
\multicolumn{1}{|l|}{$(1,2)$} & \multicolumn{1}{l|}{} & \multicolumn{1}{c|}{$(1,2,2)$} & \multicolumn{1}{c|}{}                       & \multicolumn{1}{c|}{}                   & \multicolumn{1}{c|}{}                   \\ \cline{1-1} \cline{3-3}
\multicolumn{1}{|l|}{$(2,1)$} & \multicolumn{1}{l|}{} & \multicolumn{1}{c|}{$(1,3,2)$} & \multicolumn{1}{c|}{}                       & \multicolumn{1}{c|}{}                   & \multicolumn{1}{c|}{}                   \\ \cline{1-1} \cline{3-6} 
                            & \multicolumn{1}{l|}{} & \multicolumn{1}{c|}{$(2,1,2)$} & \multicolumn{1}{c|}{\multirow{2}{*}{$(2,2)$}} & \multicolumn{1}{c|}{\multirow{2}{*}{$2$}} & \multicolumn{1}{c|}{\multirow{2}{*}{$2$}} \\ \cline{3-3}
                            & \multicolumn{1}{l|}{} & \multicolumn{1}{c|}{$(2,2,2)$} & \multicolumn{1}{c|}{}                       & \multicolumn{1}{c|}{}                   & \multicolumn{1}{c|}{}                   \\ \cline{3-6} 
                            & \multicolumn{1}{l|}{} & \multicolumn{1}{c|}{$(2,1,3)$} & \multicolumn{1}{c|}{$(2,3)$}                  & \multicolumn{1}{c|}{$1$}                  & \multicolumn{1}{c|}{$1$}                  \\ \cline{3-6} 
\end{tabular}
    \caption{\binyang{An illustration of grouping optimization}}
    \label{fig:grouping-example}
\end{figure}

\begin{example}
    \binyang{
    Consider the query $\Q = R_a(X,Y) \Join R_b(Y,Z,W) \Join R_c(W,U)$. Let $\T$ be the join tree rooted at $R_c$. We group tuples in $R_b$ by attributes $\{Y,W\}$, creating $R_{\bar{b}}(Y,W)$ as in Figure~\ref{fig:grouping-example}. The columns $\mul$ and $\wmul$ represent the values of $\mul[\T,\bar{b},t]$ and $\wmul[\T,\bar{b},t]$ respectively. By definition, $\cnt[\T, \bar{b}, (W:2)]$ equals to 10 . Suppose tuple $(3,1)$ is inserted into $R_a$. Then $\wcnt[\T, a, (Y:1)]$ increases from $2$ to $4$. Next, $\cnt[\T, \bar{b}, (W:2)]$ increases to $18$, and $\wcnt[\T, \bar{b}, (W:2)]$ increases to $32$. Note that with grouping, the change propagates through only one group $(1,2)$ instead of 3 tuples(i.e., $(1,1,2)$, $(1,2,2)$, and $(1,3,2)$).
    }
\end{example}

\paragraph{Foreign-keys} When foreign-key join exists, similar to \cite{zhao2020efficient}, we simply combine the corresponding sub-join as a whole relation. More specifically, for $R_i \Join_X R_j$, where $X$ is the primary key of $R_j$, we combine $R_i, R_j$ together as a new relation $R_{ij}= R_i \Join R_j$. This combination can be recursively done until no more foreign-key join exists. When a tuple $t_i$ is inserted into $R_i$, we check if there exists a matching tuple $t_j \in R_j$ with the value $\pi_{X} t_i$. If $t_j$ exists, we insert $t_{ij} = t_i \Join t_j$ into $R_{ij}$. However, when a tuple $t_j$ is inserted into $R_j$, we need to identify all tuples in $R_i$ that can joined with $t_j$, and insert $t_{ij} = t_i \Join t_j$ into $R_{ij}$. 

\begin{example}
Considering an example of foreign-key joins, where attributes with underscores are the primary keys of relations):
\begin{align*}
 \Q &:= R_1(\underline{X},Y) \Join R_2(Y,\underline{Z}) \Join R_3(Z,\underline{W},U) \Join R_4(\underline{U},A) \Join R_5(A,\underline{C}) \Join R_6(C,\underline{E})
 \end{align*}
After applying the combination technique, we can rewrite $\Q$ as $R_1(\underline{X},Y) \Join S(Y,Z,\underline{W},U,A) \Join T(A,C,\underline{E})$, where $S= R_2(Y,\underline{Z}) \Join R_3(Z,\underline{W},U) \Join R_4(\underline{U},A)$ and $T = R_5(A,\underline{C}) \Join R_6(C,\underline{E})$.
\end{example}

\section{Extension to Cyclic Joins}
\label{sec:cyclic}
\begin{figure}[t]   
     \resizebox{0.35\linewidth}{!}{   
     \begin{tikzpicture}
    \node (v1) at (-3,1) {};
    \node (v2) at (-3,-1) {};
    \node (v3) at (-1,0) {};
    \node (v4) at (1,0) {};
    \node (v5) at (3,1) {};
    \node (v6) at (3,-1) {};
    \node (w1) at (-3, 0) {};
    \node (w2) at (-2, 0.5) {};
    \node (w3) at (-2, -0.5) {};
    \node (w4) at (3, 0) {};
    \node (w5) at (2, 0.5) {};
    \node (w6) at (2, -0.5) {};
    \node (l1) at (-1, 1) {};
    \node (r1) at (1, 1) {};

    \begin{scope}[fill opacity=0.8]
    \filldraw[fill=yellow!30, rotate=90] ($(w4)$) ellipse (1.5 and 0.55);
    \filldraw[fill=purple!30, rotate=30] ($(w5)$) ellipse (1.5 and 0.55);
    \filldraw[fill=orange!30, rotate=150] ($(w6)$) ellipse (1.5 and 0.55);
    \filldraw[fill=blue!30, rotate=90] ($(w1)$) ellipse (1.5 and 0.55);
    \filldraw[fill=red!30, rotate=150] ($(w2)$) ellipse (1.5 and 0.55);
    \filldraw[fill=green!30, rotate=30] ($(w3)$) ellipse (1.5 and 0.55);
    \filldraw[fill=brown!70] ($(0, 0)$) ellipse (1.5 and 0.45);
    \draw[color=red, thick] ($(-2.2, 0)$) circle (1.5);
    \draw[color=red, thick] ($(2.2, 0)$) circle (1.5);
    \draw[color=red, thick] ($(0, 0)$) circle (1.45);
    \end{scope}

    \foreach \v in {1,2,...,6} {
        \fill (v\v) circle (0.1);
    }

    \fill (v1) circle (0.1) node [right] {$x_1$};
    \fill (v2) circle (0.1) node [right] {$x_2$};
    \fill (v3) circle (0.1) node [left] {$x_3$};
    \fill (v4) circle (0.1) node [right] {$x_4$};
    \fill (v5) circle (0.1) node [left] {$x_5$};
    \fill (v6) circle (0.1) node [left] {$x_6$};

    \node at (w1) {$R_1$};
    \node at (w2) {$R_2$};
    \node at (w3) {$R_3$};
    \node at (w4) {$R_4$};
    \node at (w5) {$R_5$};
    \node at (w6) {$R_6$};
    \node at (0, 0) {$R_7$};

\end{tikzpicture}
     }
    \caption{The dumbbell join $\Q = R_1(x_1,x_2)\Join R_2(x_1,x_3) \Join R_3(x_2, x_3) \Join R_4(x_5,x_6) \Join R_5(x_4,x_5) \Join R_6(x_4,x_6) \Join R_7(x_3,x_4)$ with GHD illustrated as the red circle. It has fractional hypertree width $\textsf{w}(\Q)=1.5$ since the triangle join $R_1(x_1,x_2)\Join R_2(x_1,x_3) \Join R_3(x_2, x_3)$ and $R_5(x_4,x_5) \Join R_6(x_4,x_6) \Join R_7(x_3,x_4)$ have the fractional edge covering number $\rho^*=1.5$.}
    \label{fig:dumbbell}
 \end{figure}
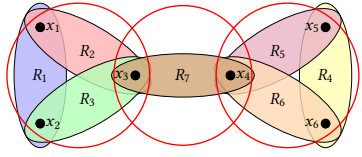

 We next show how to handle cyclic joins using our algorithm in Section~\ref{sec:acyclic}, by resorting to the classic GHD decomposition framework~\cite{gottlob2014treewidth}.
 It has been shown \cite{atserias2008size} that for a join query $\Q=(\V,\E)$ and any instance of input size $N$, the maximum join size is $\Theta(N^{\rho^*(\Q)})$, where $\rho^*(\Q)$ is the fractional edge covering number of $\Q$. \xiao{Please see an example in Figure~\ref{fig:dumbbell}.}

 \begin{definition}[Fractional Edge Covering Number]
     Given a join query $\Q = (\V,\E)$, a fractional edge covering is a function $W: \E \to [0,1]$ such that $\sum_{e\in\E: x\in e} W(e) \ge 1$ for every attribute $x \in \V$. The fractional edge covering number $\rho(\Q)$ is defined as minimum value of $\sum_{e \in \E}W(e)$ over all possible fractional edge coverings $W$.
 \end{definition}

 \begin{definition}[Generalized Hypertree Decomposition] 
    Given a join query $\Q = (\V, \E)$, a GHD of $\Q$ is a pair $(\T, \lambda)$, where $\T$ is a tree as an ordered set of nodes and $\lambda: \T \to 2^{\V}$ is a labeling function, which associates to each vertex $u \in \T$ a subset of attributes in $\V$, such that the following conditions are satisfied:
    \begin{itemize}[leftmargin=*]
        \item For each $e \in \E$, there is a node $u \in \T$ such that $e \subseteq \lambda(u)$;
        \item For each $x \in \V$, the set of nodes $\{u \in \T: x \in \lambda(u)\}$ forms a connected subtree of $\T$.
    \end{itemize}   
 \end{definition}

 Given a join query $\Q = (\V,\E)$, a GHD $(\T, \lambda)$ and a node $u \in \T$, the width of $\T$ is defined as the optimal fractional edge covering number of its derived subquery $\Q_u = (\lambda_u, \E_u)$, where $\E_u = \{e \cap \lambda_u:e \in \E\}$. Given a join query and a GHD $(\T, \lambda)$, the width of $(\T, \lambda)$ is defined as the maximum width over all nodes in $\T$. Then, the fractional hypertree width of a join query follows:
 \begin{definition}[Fractional Hypertree Width~\cite{gottlob2014treewidth}].
    The fractional hypertree width of a join query $\Q$, denoted as $\textsf{w}(\Q)$, is $\displaystyle{\textsf{w}(\Q) = \min_{(\T,\lambda)} \max_{u \in \T} \rho^*(\Q_u)}$,
    i.e., the minimum width over all GHDs.
    \end{definition}
 
 Our algorithm builds upon a GHD $(\T, \lambda)$ for the input join query $\Q$, and considers every version of $\T$ by rooting $\T$ at one distinct node. 
 Given an instance $\R$, we define the sub-instance for each node $u \in \T$ as $\R_u = \{\pi_{e \cap \lambda_u} R_e: e \in \E, e\cap \lambda_u \neq \emptyset\}$. Suppose a tuple $t$ is inserted into relation $R_e$. For each node $u \in \T$ with $e \cap \lambda_u \neq \emptyset$, we add tuple $\pi_{e \cap \lambda_u} t$ to relation $\pi_{e \cap \lambda_u} R_e$ in instance $\R_u$ and update the results of $\Q_u$. We pick an arbitrary node $u \in \T$ with $e \subseteq \lambda_u$. Let $\Delta_u = \Q_u(\R_u) \ltimes t$ be the delta join results of $t$. \xiao{For every tuple $t' \in \Delta_u$, we just execute line 5-7 in Algorithm~\ref{alg:reservoir-sampling-join}.}

 The correctness follows the fact that $\Q(\R) \ltimes t = \bigcupplus_{t' \in \Delta_u} \Q(\R) \ltimes t'$, where $\bigcupplus$ is the disjoint union operator. 
 
 \paragraph{Time Complexity} We next analyze the time complexity. For each tuple $t$ inserted into $R_e$, the join result of $\Q_u$ for each node $u$ can be updated in $\textsf{AGM}(\Q_u, \R_u \ltimes t)$ time. Summing over all inserted tuples and nodes in $\T$, the time complexity is 
 \[\sum_{u \in \T} \sum_{e \in \E}\sum_{t \in R_e} \textsf{AGM}(\Q_u, \R_u \ltimes t) \le \textsf{AGM}(\Q_u, \R_u) = O\left(N^{\textsf{w}(\Q)}\right).\]   
 Moreover, we can also bound the size of $\Delta_u(t)$ by $\textsf{AGM}(\Q_u, \R_u \ltimes t)$. Hence, the size of the simulated input stream over the GHD $(\T,\lambda)$ is bounded by $\displaystyle{\sum_{e \in \E} \sum_{t \in R_e}\textsf{AGM}(\Q_u, \R_u \ltimes t) = O\left(N^{\textsf{w}(\Q)}\right)}$. 
 
 \begin{figure*}
\includegraphics[width=0.9\linewidth]{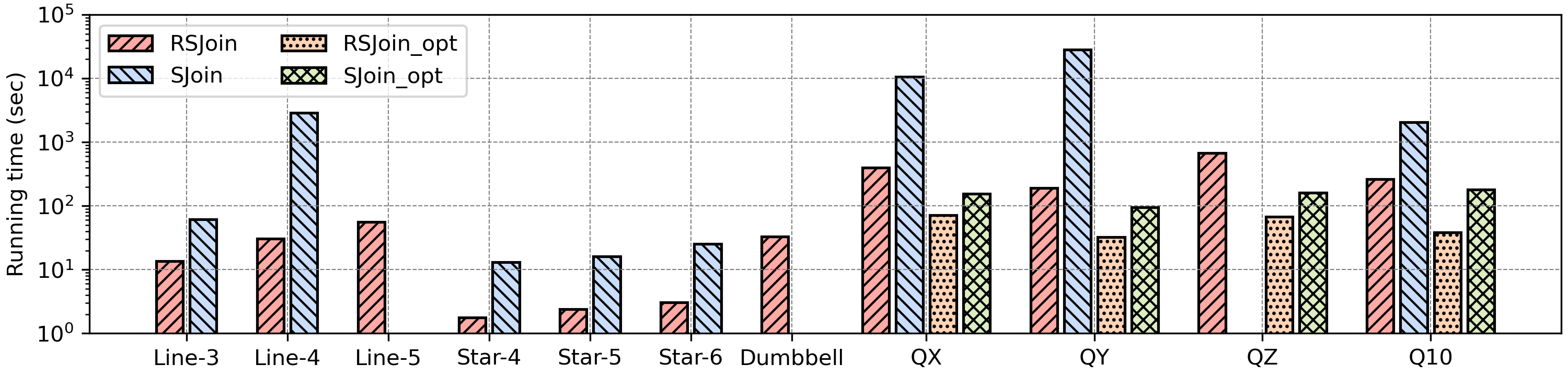}
\vspace{-1em}
   \caption{Running time over different join queries}
   \label{fig:runningtime}
 \end{figure*}

 \xiao{\paragraph{Space Usage} Our index for cyclic joins builds upon a GHD $(\T, \lambda)$ for $\Q$. Note that the total number of input tuples inserted into each node of $\T$ is $O(N^{\textsf{w}})$, where $\textsf{w}$ is the fractional hypertree width of $\Q$. Following the same analysis of acyclic joins in Section~\ref{sec:acyclic}, the space used by our index is proportional to the total number of tuples in each node of $\T$, i.e., $O(N^{\textsf{w}})$ for cyclic joins. } 
 
 \smallskip 
 Putting everything together, we obtain:
    
 \begin{theorem}
 \label{the:cyclic}
    Given any join $\Q$, an initially empty database $\R$, a sample size $k$, and a stream of $N$ tuples, \binyang{Algorithm \ref{alg:reservoir-sampling-join}} maintains $k$ uniform samples without replacement for each $\Q(\R^i)$, \xiao{uses $O(N^{\textsf{w}})$ space} and runs in $O(N^{\textsf{w}} \cdot \log N + k \cdot \log N \cdot \log \frac{N}{k})$ expected time, where $w$ is the fractional hypertree width of $\Q$.
 \end{theorem}
\section{Experiments}
\label{sec:experiment}
    \subsection{Setup} 
    \noindent {\bf Implementation.} We compare our algorithm (denoted as \batchres) as well as the optimized version when foreign-key join exists (denoted as \batchresfk), with the algorithm in~\cite{zhao2020efficient} (denoted as \sjoin) and its optimized version when foreign-key join exists (denoted as \sjoinfk), which is also the state-of-the-art method for supporting random sampling over joins under updates. \xiao{We mentioned that the symmetric hash join algorithm~\cite{SymmetricHashJoin} was proposed for computing the (delta) join results for the basic two-table join over data streams. In~\cite{zhao2020efficient}, symmetric join was combined with reservoir sampling for supporting maintaining uniform samples over joins and also tested as a baseline solution, but its performance is overall dominated by~\cite{zhao2020efficient}, hence we do not include it in our experiments.} 
    We implement our algorithms in C++, and conduct experiments on a machine equipped with two Intel Xeon 2.1GHz processors with 24 cores and 251 GB of memory, running CentOS 7. We repeat each experiment 10 times (with timeout as 12 hours) and report the average running time. All code is available at~\cite{code}.
    
    \paragraph{Datasets and Queries} We evaluate algorithms on graph and relational datasets/queries. All queries in SQL can be found in Appendix~\ref{appendix:queries}.
    
    We use the Epinions dataset that contains 508,837 edges from SNAP (Stanford Network Analysis Project)~\cite{SNAP} as the graph dataset. Each relation contains all edges. We randomly shuffle all edges for each relation to simulate the input stream. On Epinions, we evaluate line-$k$ joins (which find paths in the graph of length $k$), star-$k$ joins (which find all combinations of $k$ edges sharing a common vertex), and dumbbell join (which finds all pairs of triangle that is connected by an edge). There is no foreign-key join in graph queries.

    We use two relational datasets. One is TPC-DS dataset~\cite{tpc-ds}, which models several generally applicable aspects of a decision support system. We evaluate the same \Qx, \Qy, and \Qz\ queries as ~\cite{zhao2020efficient} on TPC-DS, which include the foreign-key joins, and follow the same setup as ~\cite{zhao2020efficient}, such that small dimension tables (such as \textsf{date\textunderscore dim} and \textsf{household\textunderscore demographics} are pre-loaded, while the rest of the tables are loaded in a streaming fashion. The other is LDBC Social Network Benchmark (LDBC-SNB) ~\cite{LDBC-SNB}, which focuses on join-heavy complex queries with updates. We tested \QSNB\ query from the Business Intelligence (BI) workload 10. Similar as before, the static tables (such as \textsf{tag} and \textsf{city}) are pre-loaded, and the dynamic tables are loaded in a streaming fashion.

 \subsection{Experiment Results}
 \noindent {\bf Running time.}
 Figure~\ref{fig:runningtime} shows the running time of all algorithms on tested queries. For graph queries (i.e., line-$k$, star-$k$, and dumbbell),the sample size is 100,000. For relational queries (i.e., \Qx, \Qy, \Qz, and \QSNB), the sample size is 1,000,000. For the TPC-DS dataset, we use a scale factor of 10, while for the LDBC-SNB dataset, we use a scale factor of 1. Firstly, \batchres\ and \batchresfk\ can finish all queries within 12-hour time limit while \sjoin\ cannot finish on the line-5 join and the \Qz\ join. For the dumbbell join, the result is missing for \sjoin\ since it does not support cyclic queries. Secondly, \batchres\ is always the fastest over all join queries. Based on existing results, \batchres\ achieves a speedup ranging from 4.6x to 147.6x over \sjoin, not mention the case when \sjoin\ cannot finish in time. When foreign-key join exists (i.e., \Qx, \Qy, \Qz, and \QSNB), \batchresfk\ achieves an improvement of 2.2x to 4.7x over \sjoinfk. Furthermore, for \Qx, \Qy, \Qz, and \QSNB, \batchres\ does not heavily rely on foreign-key optimizations as \sjoin. As long as data satisfies foreign-key constraints, \batchres\ finishes the execution within a reasonable amount of time, but this is not the case for \sjoin. 

\begin{figure*}
  \centering
  \begin{minipage}{0.48\linewidth}
  \includegraphics[scale=0.5]{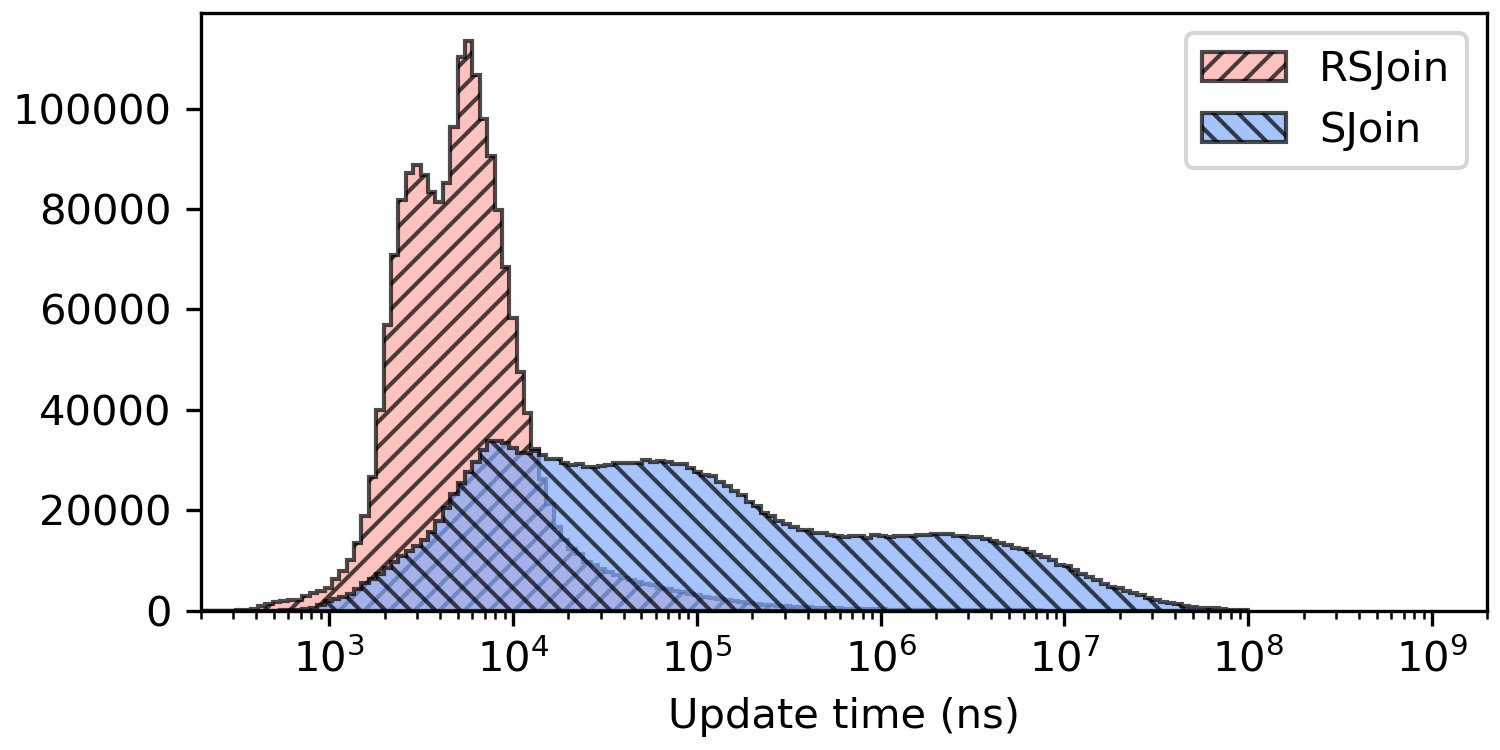}
  \vspace{-1em}
  \caption{Update time distribution}
  \label{fig:updatetime}
  \end{minipage}
  \begin{minipage}{0.48\linewidth}
  \includegraphics[scale=0.5]{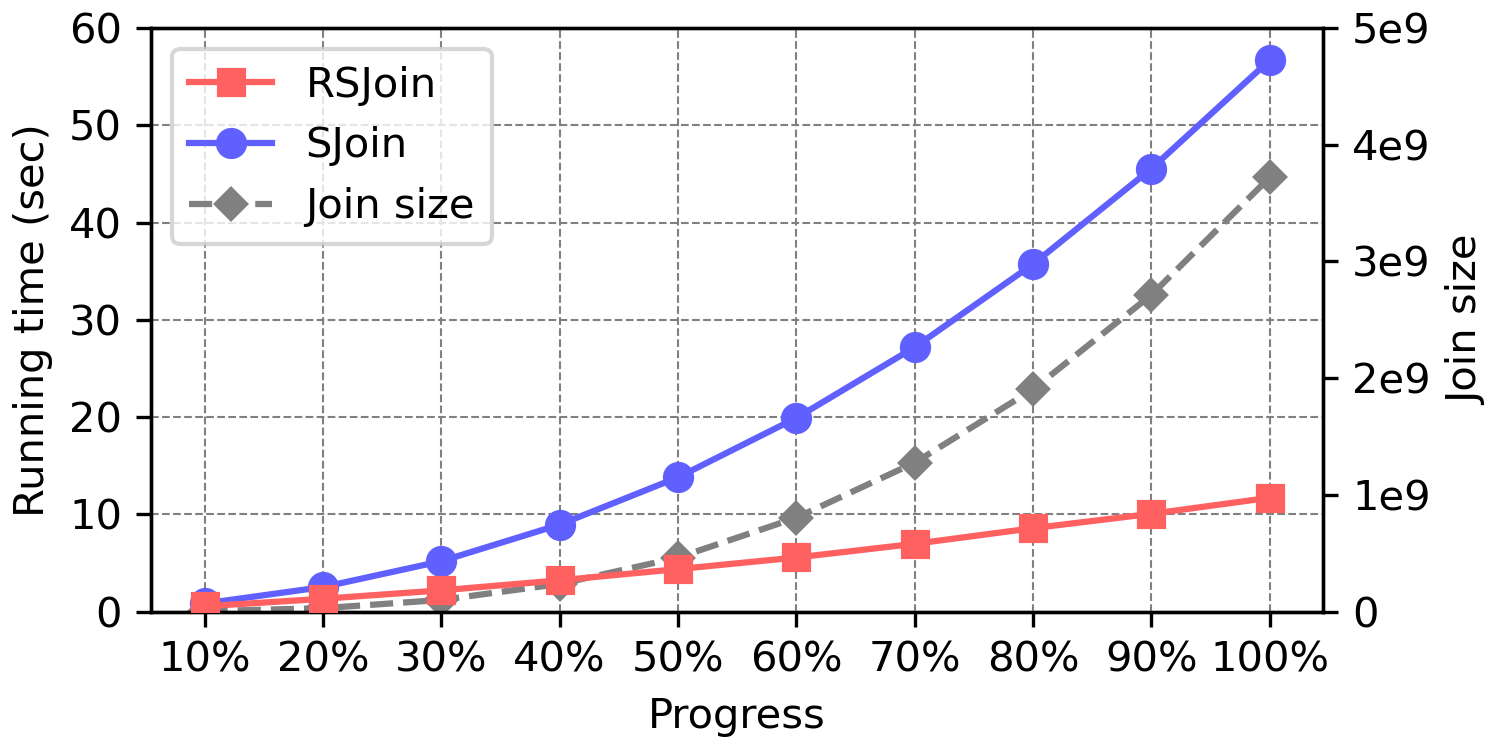}
  \vspace{-1em}
  \caption{Running time v.s. input size and join size}
  \label{fig:input_size}
  \end{minipage}
 \end{figure*}

 \smallskip \noindent{\bf Update time.} 
 To compare the update time, we disable the sampling part of both algorithms and measure the update time required for each input tuple. Figure~\ref{fig:updatetime} shows the result on line-4 join. Most of the update time required is roughly 10 \textmu s, with an average of 13 \textmu s. Some tuple may incur much larger update time(51 ms in this case), but the overall update time remains small, which aligns with our theoretical analysis of $O(\log N)$ amortized update time. In contrast, there is no guarantee on the update time for \sjoin, and its update time ranges from 0.5 \textmu s to 165 ms, with an average of 1.4 ms.

 \paragraph{Input size and Join size} We next investigate how the input size $N$ as well as the join size (i.e., the number of join results) affect the total execution time of all methods. We fix the sample size $k$ to be $10,000$ and record the total execution after every 10\% of input data is processed for line-3 join. 
 Figure~\ref{fig:input_size} shows the progress of total number of join results generated and the total execution time. We can see that the total number of join results grows exponentially with the input size, while the total execution time of \batchres\ scales almost linearly proportional to the input size, instead of the join size. This is expected as the time complexity of \batchres\ is $O(N \cdot  \log N + k \cdot \log N \cdot \log \frac{N}{k})$, where the term $N\log N$ almost dominates the total execution time in this case. In contrast, the total execution time of \sjoin\ shows a clear increase trend together with the increase in the join size, which is much larger than the input size.  

 \begin{figure*}
  \centering
  \begin{minipage}{0.48\linewidth}
  \includegraphics[scale=0.5]{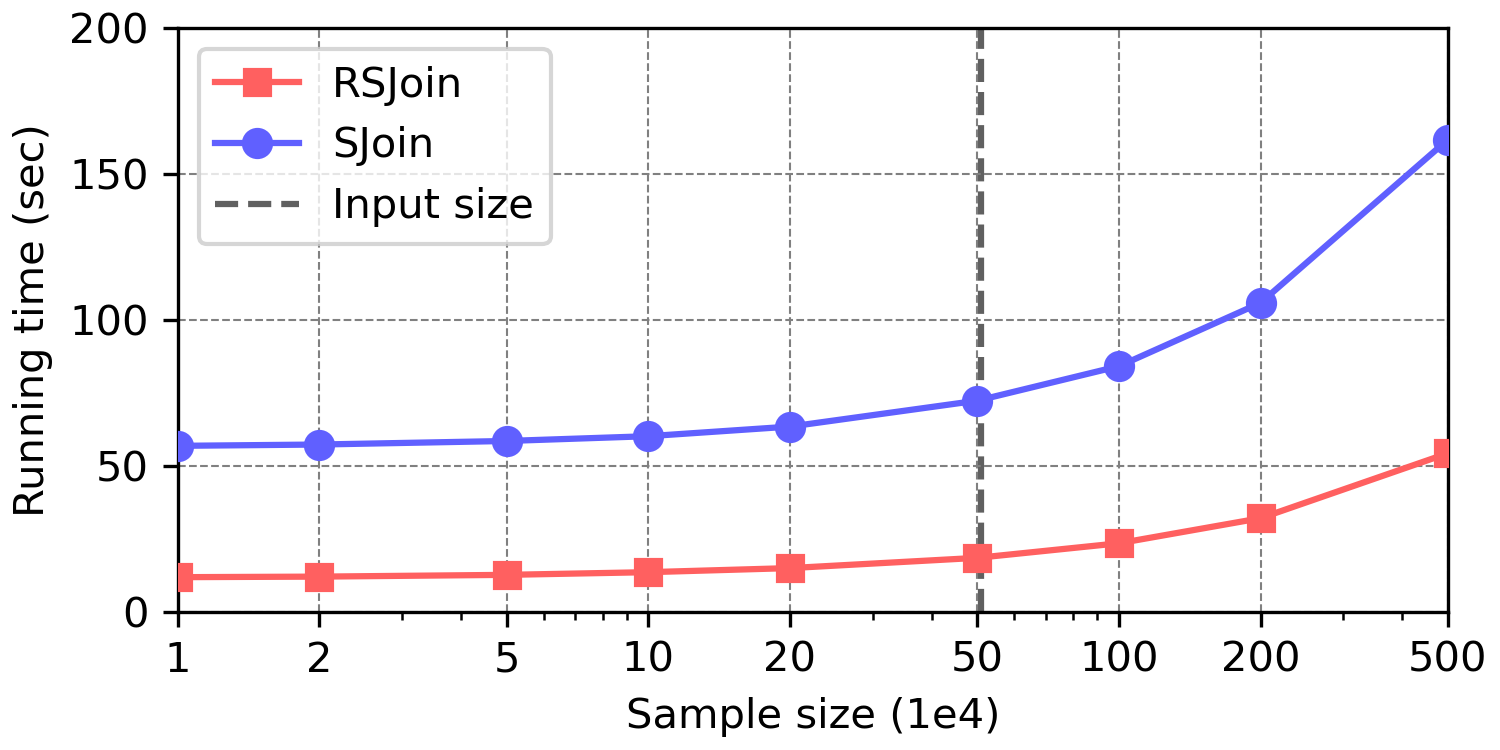}
  \vspace{-1em}
  \caption{Running time v.s. sample size}
  \label{fig:sample_size}
  \end{minipage}
  \begin{minipage}{0.48\linewidth}
    \centering
     \vspace{2.6em}
    {\small
    \begin{tabular}{||c|c|c||} 
    \hline
    Optimizations & \#Execution & Run-time (sec) \\ 
    \hline
    N/A & 172010370 & 678.864 \\ 
    \hline
    Foreign-key & 132175648 & 204.614 \\ 
    \hline
    Foreign-key +  & \multirow{2}{*}{597557} & \multirow{2}{*}{68.047} \\
    Grouping & & \\
    \hline
    \end{tabular}
    \caption{Optimizations on \Qz\ over TPC-DS dataset}
    \label{table:foreign_key}
    }
  \end{minipage}
 \end{figure*}

 \paragraph{Sample size} We next study how the sample size $k$ affect the total execution time of both algorithms. Figure~\ref{fig:sample_size} shows the running time on line-3 join, when $k$ varies from 10,000 to 5,000,000. The dashed line indicates the input size $N = 508,837$ and the number of join results is 3,721,042,797. When the sample size is smaller than the input size, i.e., $k \le N$, the total execution time of \batchres\ grows very slow. More specially, when $k$ increases from $1$ to $50$, the total execution time of \batchres\ only increases by a factor of $2$. However, when the sample time overrides the input size, i.e., $k > N$, the total execution time of \batchres\ starts to increases rapidly. This is also expected again as the theoretical complexity of \batchres\ is $O(N \cdot \log N + k \cdot \log N \cdot \log \frac{N}{k})$. When $k \le N$, the term $O(N \cdot \log N)$ dominates the overall execution time, hence increasing the sample size within this regime does not change the total execution time significantly.  When $k > N$, the term $O(k \cdot \log N \cdot \log{\frac{N}{k}})$ dominates the overall execution time instead, hence increasing the sample size results in rapid increase in the total execution time. \sjoin\ follows a similar trend. Moreover, when the sample size reaches $k=10,000$, the running time required by \sjoin\ is even more than that required by \batchres\ for the case when  sample size is as large as $k=5,000,000$. 

\begin{figure*}
  \centering
  \begin{minipage}{0.48\linewidth}
  \includegraphics[scale=0.5]{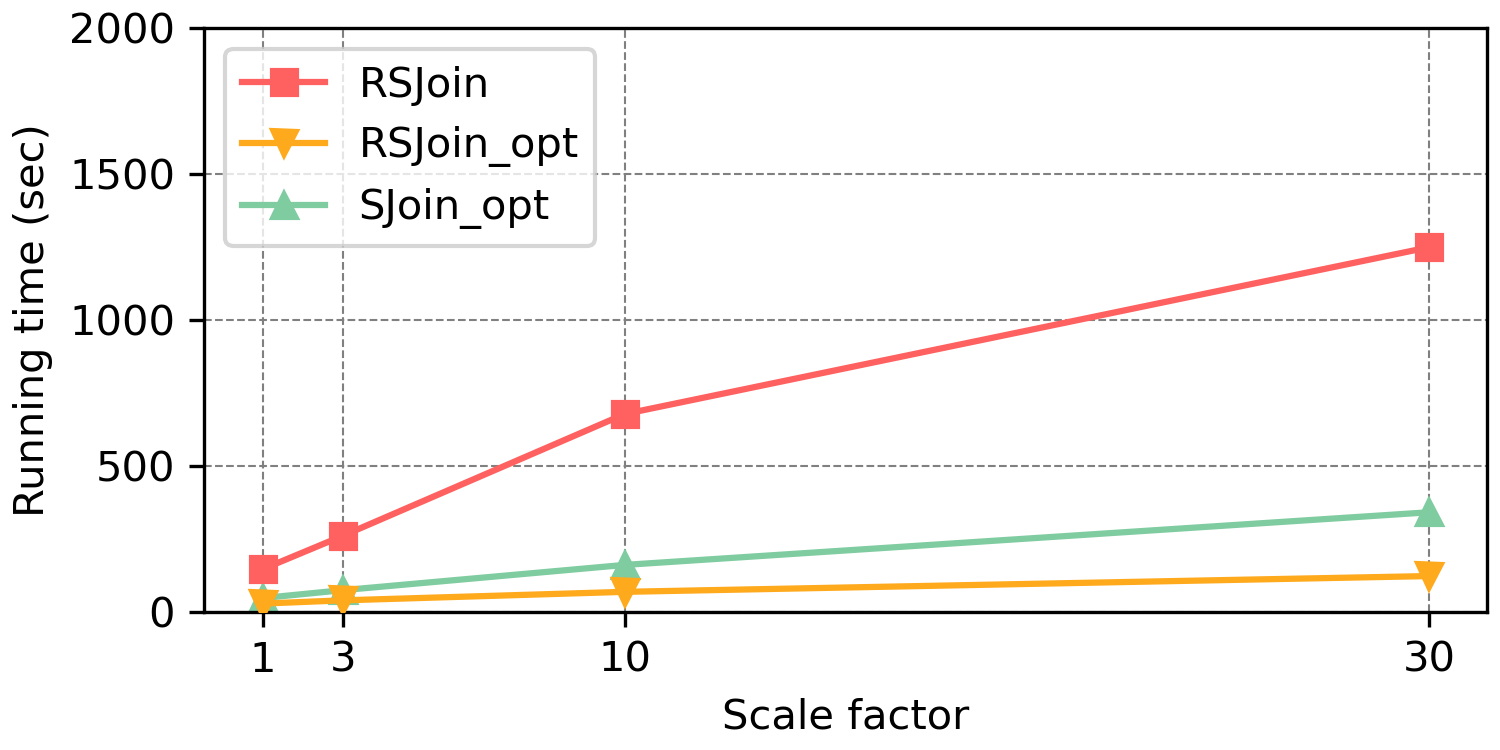}
  \vspace{-1em}
  \caption{Running time v.s. scale factor}
  \label{fig:scalability}
  \end{minipage} \ \ \ 
  \begin{minipage}{0.48\linewidth}
  \includegraphics[scale=0.5]{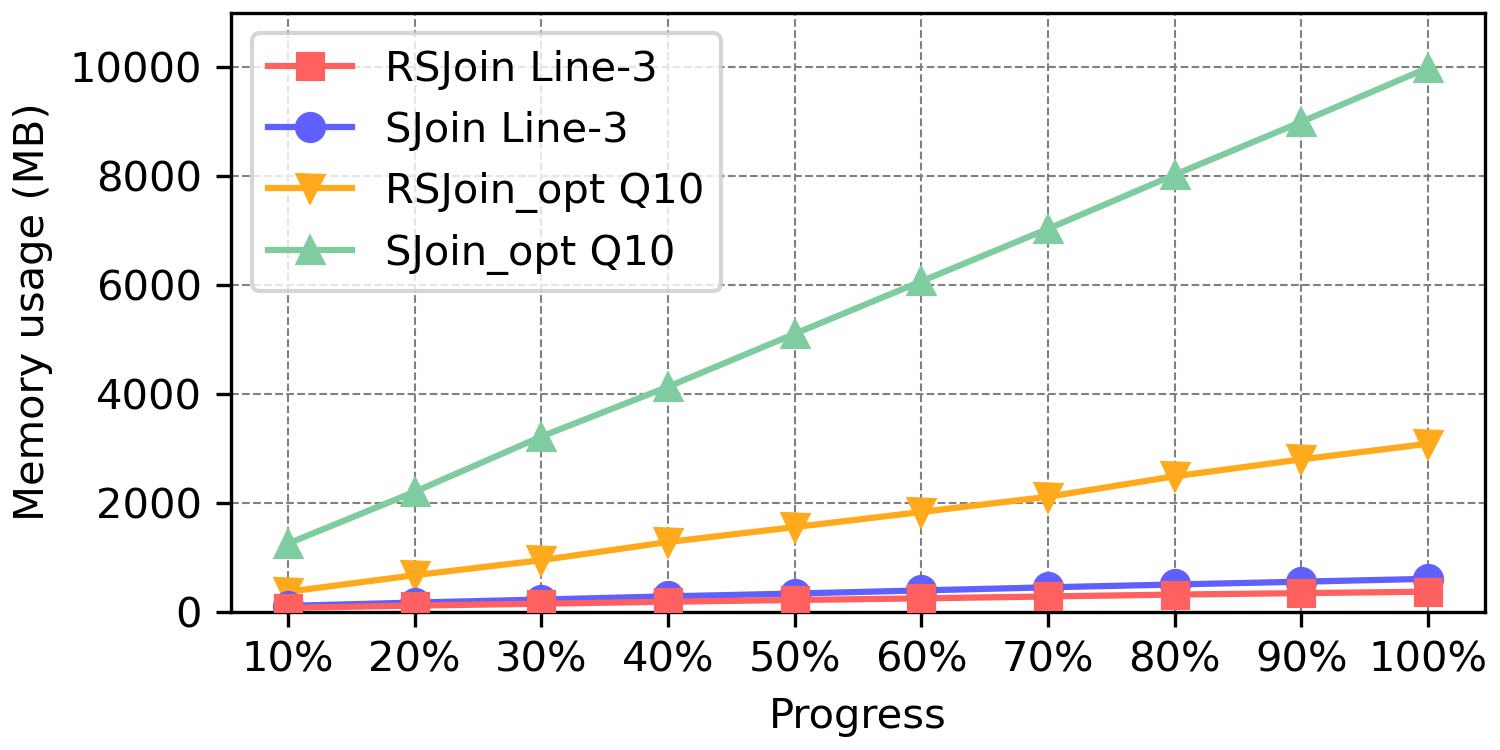}
  \vspace{-1em}
  \caption{Memory usage v.s. input size}
  \label{fig:memory_usage}
  \end{minipage}
 \end{figure*}

 \paragraph{Scalability} To examine the scalability of both methods, we evaluate the \Qz\ query on TPC-DS dataset with scale factors of 1,3,10, and 30. The results are shown in Figure~\ref{fig:scalability}. The input size of \Qz\ is approximately 226MB when the scale factor is 1, while the input size reaches around 6.6GB when the scale factor reaches 30. We do not include the results of \sjoin\ here since it takes more than 4 hours to finish the execution even with scale factor as 1. We observe that even without applying foreign-key optimization, \batchres\ achieves linear growth in the running time as the scale factor increases, which indicates that \batchres\ is scalable and practical even when dealing with significantly huge input size.
  \begin{figure*}
   \centering
   \begin{minipage}{0.48\linewidth}
   \includegraphics[width=0.9\linewidth]{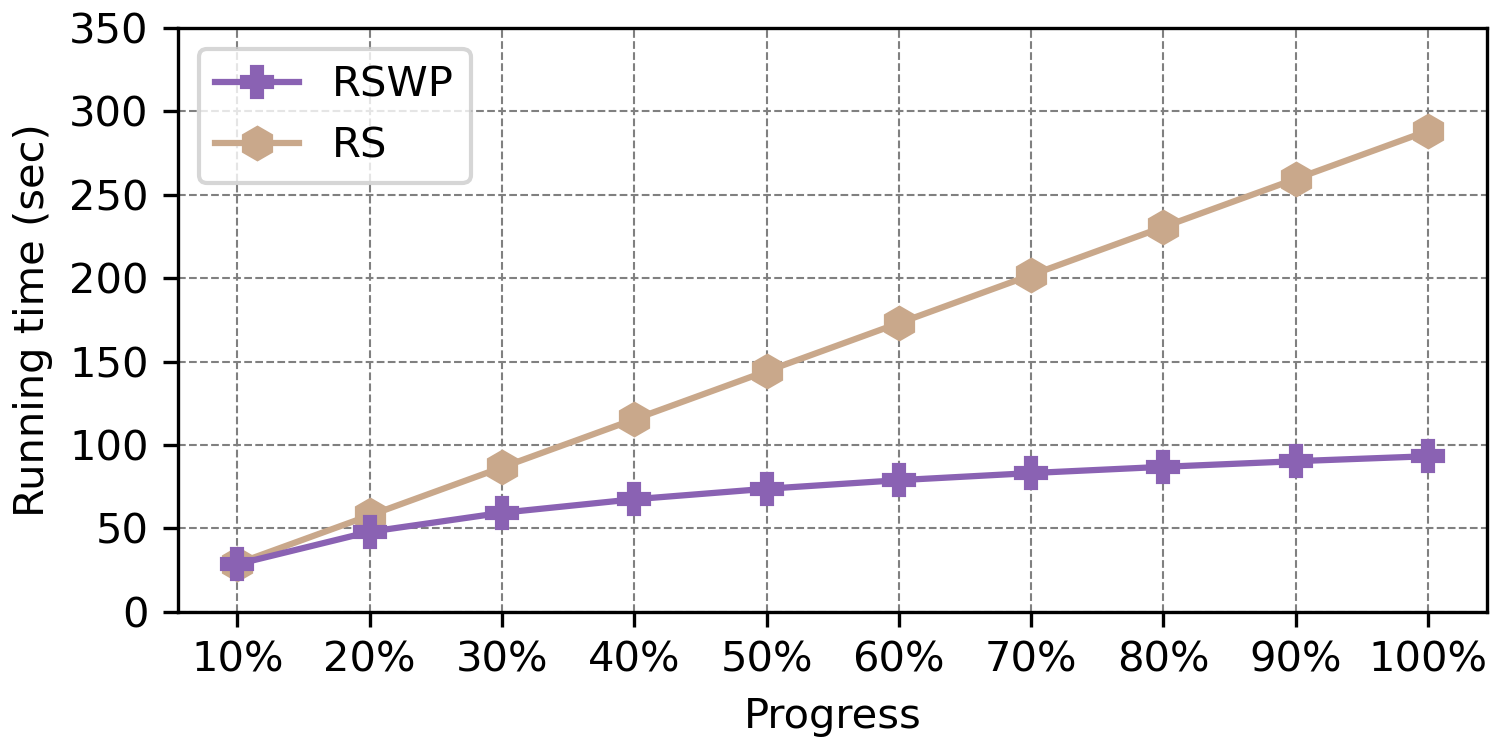}
   \vspace{-1em}
   \caption{\xiao{Running time v.s. input size}}
   \label{fig:reservior-fixed-density}
   \end{minipage}
   \begin{minipage}{0.48\linewidth}
   \includegraphics[width=0.9\linewidth]{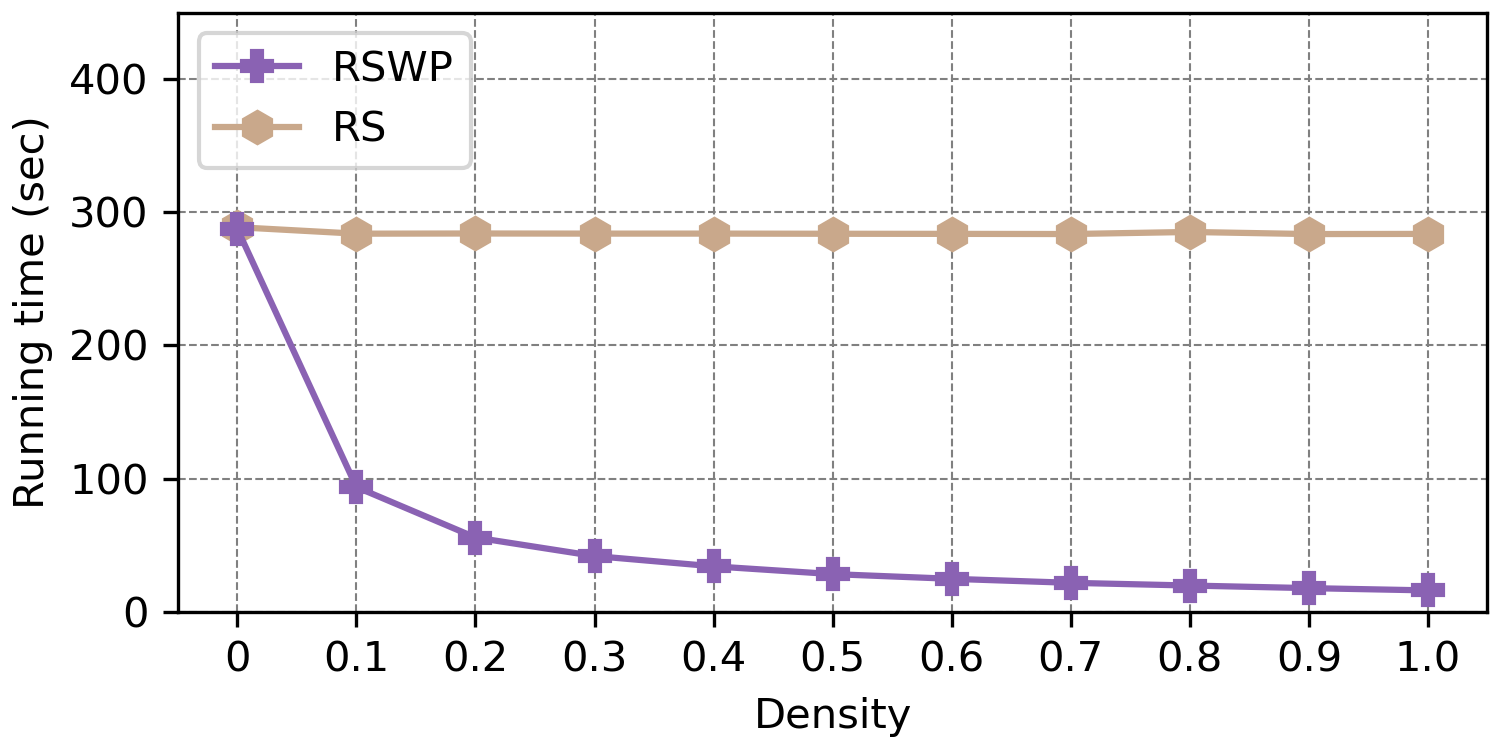}
   \vspace{-1em}
   \caption{\xiao{Running time v.s. density}}
   \label{fig:reservior-vary-density}
   \end{minipage}
   \end{figure*}

 \paragraph{Memory usage} In addition, we explore the memory usage of all methods. Figure~\ref{fig:memory_usage} shows the memory usage by \batchres\ and \sjoin\ on line-3 join and \batchresfk\ and \sjoinfk\ on \QSNB\ query. The input size is roughly 21MB for line-3 join and  505MB for \QSNB\ query. After processing every 10\% of the input data, we record the memory usage as shown in Figure~\ref{fig:memory_usage}. The memory usage of \QSNB\ grows much faster than line-3 join as it is much more complex with more dedicated index built. The memory usage required by all algorithms is linear to the input size. On line-3 join, \batchres\ requires only 60\% of the memory by \sjoin, and on \QSNB, \batchresfk\ needs only 31\% of the memory by \sjoinfk. This demonstrates a nice property of our algorithm: the amount of memory used by \batchres\ and \batchresfk\ during execution scales linearly with the input size even when the join size grows exponentially, which also enables our algorithm to handle much more complex queries over large input datasets with limited memory resources.
  
 \paragraph{Optimizations} We evaluate the effectiveness of our optimizations by \xiao{counting the number of loop execution lines 9-11 in Algorithm~\ref{alg:index-update}.} Table~\ref{table:foreign_key} records the count as well as the total running time of our method for \Qz\ query over the TPC-DS dataset (scale factor 10 and sample size 1,000,000). It is clear to see that when applying foreign-key optimizations, the number of propagation decreases as well as the total execution time. If applying the grouping optimizations, we can further decrease the total execution time, achieving roughly 10x speedup over the \batchres\ without optimization. 
  
  \subsection{\xiao{Reservoir Sampling with Predicate}}
  \label{sec:experiment-rs}
   \binyang{At last, we compare our new reservoir sampling algorithm with predicate (denoted as \textsf{RSWP}) with the classic reservoir sampling algorithm (denoted as \textsf{RS}) on data streams. We generate a data stream as follows.  We fix a random string of 1024 characters, referred as the query string. Each item in the input stream is a random string, within edit distance ranging from 0 to 64 from the base string. The predicate selects all strings in the data stream whose edit distance from the query string is less than or equal to 16. 
   
   In Figure~\ref{fig:reservior-fixed-density}, we take a $\frac{1}{10}$-dense stream of $100,000$ strings with sample size $k=1,000$. 
   We record the execution time after processing every $10\%$ of the input stream. As \textsf{RS} needs to process every item (i.e., compute the edit distance from the query string), the running time of \textsf{RS} is linear to the number of items in the stream processed so far. 
   The time required by \textsf{RSWP} for processing the first $10\%$ of the input stream is the same as \textsf{RS}, since both of them need to process every one in the first 10,000 items (approximately) until it fills the reservoir. After that, the running time of \textsf{RSWP} grows slower and slower, which is consistent with our theoretical result that it takes $O\big(\frac{k}{r_i + 1}\big)$ expected time to process the $i$-th item.   

   In figure~\ref{fig:reservior-vary-density}, we measure the running time of both \textsf{RSWP} and \textsf{RS} over $11$ streams of same input size but different densities. As \textsf{RS} needs to process every item in the stream, its running time only depends on the input size, instead of the density of input stream. In contrast,  the running time of \textsf{RSWP} depends on the density of stream. In an extreme case, when no item passes the predicate (i.e., the density is $0$), \textsf{RSWP} cannot skip any item and hence requires the same time as \textsf{RS}. However, as density increases, the running time of \textsf{RSWP} decreases significantly. In another extreme case, when every item passes the predicate (i.e., the density is $1.0$), \textsf{RSWP} exhibits a speed advantage of 17.7x over \textsf{RS}.}
\section{Related Work}
\label{sec:related}
In addition to the directly related work mentioned in Section \ref{sec:previous}, the following is also relevant to our work:

\paragraph{Streaming Subgraphs Sampling} The problem of sampling subgraph patterns from a graph whose edges come as an input stream has also been considered (where the space usage is important). For example, Paven et. al~\cite{pavan2013counting} designed an
algorithm that uses $O(\frac{N^{3/2}}{\OUT})$ space, where $\OUT$ is the number of triangles in the graph. In the field of property testing (where sub-linear number of
query accesses to the graph is important), Eden et al~\cite{eden2017sampling} studied the problem of almost uniform sampling of edges, and Biswas et. al~\cite{biswas2021towards} studied the problem of sampling
subgraphs.

\paragraph{Maintaining Conjunctive Queries under Updates}
It has been shown \cite{idris2017dynamic, berkholz17:_answer} that a very restrictive class of queries, known as q-hierarchical query can admit an index with $O(1)$ update time. However, any non-q-hierarchical query, a lower bound of $\Omega(N^{\frac{1}{2} - \epsilon})$ has also been proved on the update time, for any small constant $\epsilon >0$. This result is rather negative, since q-hierarchical queries are a very restricted class; for example, the line-3 join. Meanwhile, \cite{idris2017dynamic} showed an index for acyclic joins that can be updated in $O(N)$ time. Later, Kara et al.~\cite{kara2019counting} designed optimal data structures that can be updated in $O(\sqrt{N})$ time while supporting $O(1)$-delay enumeration for line-3 join, triangle join, length-4 cycle join, etc. Moreover, Kara et al.~\cite{kara2020trade} also investigated the tradeoff between update time and delay for hierarchical queries. Wang et al.~\cite{wang2020maintaining, wang2023change} worked on instance-dependent complexity by relating the update time to the enclosureness of update sequences. 
\section{Conclusion}
\label{sec:conclusion}

In this paper, we propose a general reservoir sampling algorithm that supports a predicate. We design a dynamic data structure that supports efficient updates and direct access of the join results. By combining these two key techniques, we present our reservoir sampling over joins algorithm which runs in near-linear time. There are several interesting questions left as open, such as uniform sampling over join-project queries over data streams.

\bibliographystyle{ACM-Reference-Format}
\bibliography{paper_full}

\appendix
\section{Queries}
\label{appendix:queries}
\paragraph{Line-3 Join}
\lstset{style=mystyle}
\begin{lstlisting}[ language=SQL,
	deletekeywords={IDENTITY},
	deletekeywords={[2]INT},
	morekeywords={clustered},
	mathescape=true,
	xleftmargin=-1pt,
	framexleftmargin=-1pt,
	frame=tb,
	framerule=0pt ]
SELECT G1.src AS A, G2.src AS B, G3.src AS C, G3.dst AS D
FROM G AS G1, G AS G2, G AS G3 
WHERE G1.dst = G2.src AND G2.dst = G3.src 
\end{lstlisting}

\paragraph{Line-4 Join}
\lstset{style=mystyle}
\begin{lstlisting}[ language=SQL,
	deletekeywords={IDENTITY},
	deletekeywords={[2]INT},
	morekeywords={clustered},
	mathescape=true,
	xleftmargin=-1pt,
	framexleftmargin=-1pt,
	frame=tb,
	framerule=0pt ]
SELECT G1.src AS A, G2.src AS B, G3.src AS C, G4.src AS D, G4.dst AS E
FROM G AS G1, G AS G2, G AS G3, G AS G4
WHERE G1.dst = G2.src AND G2.dst = G3.src AND G3.dst = G4.src
\end{lstlisting}

\paragraph{Line-5 Join}
\lstset{style=mystyle}
\begin{lstlisting}[ language=SQL,
	deletekeywords={IDENTITY},
	deletekeywords={[2]INT},
	morekeywords={clustered},
	mathescape=true,
	xleftmargin=-1pt,
	framexleftmargin=-1pt,
	frame=tb,
	framerule=0pt ]
SELECT G1.src AS A, G2.src AS B, G3.src AS C, G4.src AS D, G5.src AS E, G5.dst AS F
FROM G AS G1, G AS G2, G AS G3, G AS G4, G AS G5
WHERE G1.dst = G2.src AND G2.dst = G3.src AND G3.dst = G4.src AND G4.dst = G5.src
\end{lstlisting}

\paragraph{Star-4 Join}
\lstset{style=mystyle}
\begin{lstlisting}[ language=SQL,
	deletekeywords={IDENTITY},
	deletekeywords={[2]INT},
	morekeywords={clustered},
	mathescape=true,
	xleftmargin=-1pt,
	framexleftmargin=-1pt,
	frame=tb,
	framerule=0pt ]
SELECT *
FROM G AS G1, G AS G2, G AS G3, G AS G4
WHERE G1.src = G2.src AND G1.src = G3.src AND G1.src = G4.src
\end{lstlisting}

\paragraph{Star-5 Join}
\lstset{style=mystyle}
\begin{lstlisting}[ language=SQL,
	deletekeywords={IDENTITY},
	deletekeywords={[2]INT},
	morekeywords={clustered},
	mathescape=true,
	xleftmargin=-1pt,
	framexleftmargin=-1pt,
	frame=tb,
	framerule=0pt ]
SELECT *
FROM G AS G1, G AS G2, G AS G3, G AS G4, G AS G5
WHERE G1.src = G2.src AND G1.src = G3.src AND G1.src = G4.src AND G1.src = G5.src
\end{lstlisting}

\paragraph{Star-6 Join}
\lstset{style=mystyle}
\begin{lstlisting}[ language=SQL,
	deletekeywords={IDENTITY},
	deletekeywords={[2]INT},
	morekeywords={clustered},
	mathescape=true,
	xleftmargin=-1pt,
	framexleftmargin=-1pt,
	frame=tb,
	framerule=0pt ]
SELECT *
FROM G AS G1, G AS G2, G AS G3, G AS G4, G AS G5, G AS G6
WHERE G1.src = G2.src AND G1.src = G3.src AND G1.src = G4.src AND G1.src = G5.src AND G1.src = G6.src
\end{lstlisting}

\paragraph{Dumbbell Join}
\lstset{style=mystyle}
\begin{lstlisting}[ language=SQL,
	deletekeywords={IDENTITY},
	deletekeywords={[2]INT},
	morekeywords={clustered},
	mathescape=true,
	xleftmargin=-1pt,
	framexleftmargin=-1pt,
	frame=tb,
	framerule=0pt ]
SELECT *
FROM G AS G1, G AS G2, G AS G3, G AS G4, G AS G5, G AS G6, G AS G7
WHERE G1.dst = G2.src AND G2.dst = G3.src AND G3.dst = G1.src 
AND G4.dst = G5.src AND G5.dst = G6.src AND G6.dst = G4.src 
AND G1.src = G7.src AND G4.src = G7.dst
\end{lstlisting}

\paragraph{\Qx\ Join}
\lstset{style=mystyle}
\begin{lstlisting}[ language=SQL,
	deletekeywords={IDENTITY},
	deletekeywords={[2]INT},
	morekeywords={clustered},
	mathescape=true,
	xleftmargin=-1pt,
	framexleftmargin=-1pt,
	frame=tb,
	framerule=0pt ]
SELECT *
FROM store_sales, store_returns, catalog_sales,
date_dim d1, date_dim d2 WHERE ss_item_sk = sr_item_sk
AND ss_ticket_number = sr_ticket_number AND sr_customer_sk = cs_bill_customer_sk AND d1.d_date_sk = ss_sold_date_sk
AND d2.d_date_sk = cs_sold_date_sk;
\end{lstlisting}

\paragraph{\Qy\ Join}
\lstset{style=mystyle}
\begin{lstlisting}[ language=SQL,
	deletekeywords={IDENTITY},
	deletekeywords={[2]INT},
	morekeywords={clustered},
	mathescape=true,
	xleftmargin=-1pt,
	framexleftmargin=-1pt,
	frame=tb,
	framerule=0pt ]
SELECT *
FROM store_sales, customer c1, household_demographics d1,
customer c2, household_demographics d2 WHERE ss_customer_sk = c1.c_customer_sk
AND c1.c_current_hdemo_sk = d1.hd_demo_sk
AND d1.hd_income_band_sk = d2.hd_income_band_sk AND d2.hd_demo_sk = c2.c_current_hdemo_sk;
\end{lstlisting}

\paragraph{\Qz\ Join}
\lstset{style=mystyle}
\begin{lstlisting}[ language=SQL,
	deletekeywords={IDENTITY},
	deletekeywords={[2]INT},
	morekeywords={clustered},
	mathescape=true,
	xleftmargin=-1pt,
	framexleftmargin=-1pt,
	frame=tb,
	framerule=0pt ]
SELECT *
FROM store_sales, customer c1, household_demographics d1,
item i1, customer c2, household_demographics d2, item i2 WHERE ss_customer_sk = c1.c_customer_sk
AND c1.c_current_hdemo_sk = d1.hd_demo_sk
AND d1.hd_income_band_sk = d2.hd_income_band_sk AND d2.hd_demo_sk = c2.c_current_hdemo_sk
AND ss_item_sk = i1.i_item_sk
AND i1.i_category_id = i2.i_category_id;
\end{lstlisting}

\paragraph{Q10 Join}
\lstset{style=mystyle}
\begin{lstlisting}[ language=SQL,
	deletekeywords={IDENTITY},
	deletekeywords={[2]INT},
	morekeywords={clustered},
	mathescape=true,
	xleftmargin=-1pt,
	framexleftmargin=-1pt,
	frame=tb,
	framerule=0pt ]
SELECT *
FROM Message, Tag AS Tag1, Tag AS Tag2, City, Country,
HasTag AS HasTag1, HasTag AS HasTag2,
TagClass, Person AS Person1, Person AS Person2, Knows
WHERE Message.id = HasTag1.message_id 
AND HasTag1.tag_id = Tag1.id 
AND Message.id = HasTag2.message_id 
AND HasTag2.tag_id = Tag2.id 
AND Tag2.type_tag_class_id = TagClass.id 
AND Message.creator_person_id = Person1.id 
AND Person1.location_city_id = City.id 
AND City.part_of_place_id = Country.id 
AND Person1.id = Knows.person1_id 
AND Knows.person2_id = Person2.id
\end{lstlisting}

\end{document}